# Role of hydrogen in decarbonizing China's electricity and hard-to-abate sectors

Haozhe Yang[1,2,*], Gang He[3], Eric Masanet[1,4], Ranjit Deshmukh[1,5,*]

[1] Bren School of Environmental Science and Management, University of California Santa Barbara, Santa Barbara, California, 93106, USA

[2] Andlinger Center for Energy and Environment, Princeton University, Princeton, New Jersey, 08540, USA

[3] Marxe School of Public and International Affairs, Baruch College, City University of New York, New York, 10010, USA

[4] Department of Mechanical Engineering, University of California Santa Barbara, CA, 93106, USA

[5] Environmental Studies Program, University of California Santa Barbara, Santa Barbara, California, 93106, USA

Correspondence: haozheyang@princeton.edu, rdeshmukh@ucsb.edu

**Abstract**

Green hydrogen has the potential to address two pressing problems in a zero-carbon energy system: balancing seasonal variability of solar and wind in the electricity sector, and replacing fossil fuels in hard-to-abate sectors. However, the previous research only separately modeled the electricity and hard-to-abate sectors, which is unable to capture how the interaction between the two sectors influences the energy system cost. In this study, focusing on China, we deploy an electricity system planning model to examine the cost implications of green hydrogen to fully decarbonize the electricity system and hard-to-abate sectors. Our results reveal that green hydrogen enables a 17% reduction in the levelized cost of a zero-carbon electricity system relative to that without hydrogen—however, cost savings hinge on the availability of underground hydrogen storage capacities and electric transmission expansion. More importantly, coupling hydrogen infrastructure in the electricity and hard-to-abate sectors not only reduces energy costs compared to a decoupled energy system but also makes green hydrogen cost-competitive compared to fossil fuel-based gray and blue hydrogen in China.

**Keywords**: electricity, hard-to-abate sectors, hydrogen, carbon neutral, zero-emission, energy system

**Introduction**

Green hydrogen, which is produced by electrolysis of water using renewable energy, has emerged as a promising long-term energy storage technology that can help integrate high shares of wind and solar and mitigate the high costs of a zero-carbon emission electricity system.[1,2] Despite rapid declines in the costs of solar and wind energy[3], achieving a zero-carbon electricity system with only solar, wind, and battery storage increases the costs by two- or three-fold relative to a system without a carbon emissions target, which additionally relies on coal and natural gas[4]. This increase in cost is mainly caused by the lack of long-term storage to balance the diurnal and seasonal variation of solar and wind, which results in curtailment of energy generation[5–7]. Instead, when solar and wind are abundant, the oversupply of electricity, which otherwise would be curtailed, can be converted to hydrogen and stored for several weeks before being converted back to electricity when solar and wind generation is insufficient to meet demand. Previous research has shown that green hydrogen coupled with abundant underground storage sites could reduce the cost of zero-carbon electricity systems in the United States[8] and Europe[9,10] by approximately 15% compared to that without hydrogen.

In addition to serving as long-term storage in the electricity system, green hydrogen is a promising alternative fuel and/or feedstock to replace fossil fuels in hard-to-abate sectors including heavy industry (refineries, iron and steel, ammonia, and methanol) and transportation[11]. Hard-to-abate sectors rely on fossil fuels for both fuel and feedstock due to their physical, technological, and economic characteristics. These hard-to-abate sectors account for about one-fifth of global carbon emissions. While some processes in these hard-to-abate sectors can be electrified and powered by decarbonized grids, electrifying many of the processes in these sectors is economically and technically infeasible[12]. Due to its technology maturity and declining costs, green hydrogen has the potential to be a key solution to decarbonize hard-to-abate sectors[13]. Recent research found that deploying hydrogen in a net-zero energy system reduces the global mitigation cost of carbon dioxide ($CO_2$) by up to 22%[14], but used a low temporal resolution which is unable to capture the hourly variation of wind and solar.

Due to its potentially cost-effective role in a zero-carbon emission energy system, hydrogen may be key to deep decarbonization in countries with substantial greenhouse gas emissions, e.g., China[15]. China has committed to reach the peak of its carbon emissions by 2030 and achieve net zero greenhouse gas emissions by 2060[16]. To reach China's net zero target, $CO_2$ emissions from the electricity and hard-to-abate sectors need to reach zero by 2050[17]. Previous studies have separately assessed the role of hydrogen in China's electricity system[18–20] and hard-to-abate sectors[11,21,22]. However, three key questions remain unanswered.

First, what are the impacts on system costs by using hydrogen in a zero-carbon emission electricity system by only expanding the capacities of solar and wind? Specifically, what are the limiting geographical and technological factors for hydrogen development, and how does hydrogen interplay with other low-carbon technologies like nuclear, and carbon capture and storage (CCS), and direct air capture (DAC)? Without hydrogen, previous research found that China's zero-carbon electricity system would need to rely heavily on nuclear, hydropower, and carbon removal technologies. Under this plan, nuclear power capacity roughly quadruples and hydropower capacity doubles by 2050, relative to 2020 levels[23,24]. Coal, natural gas, or bioenergy power plants aggregately provide 10-20% of electricity demand, with their carbon emissions captured by CCS[23–26].

Second, what are the cost implications of coupling green hydrogen in the zero-carbon electricity system with hard-to-abate sectors? By coupling the electricity system and hard-to-abate sectors, green hydrogen from the electricity system not only serves as long-term storage, but also meets hydrogen demand in hard-to-abate sectors. When electricity and hard-to-abate sectors are separately assessed, two important mechanisms in the system cost are not considered: the antagonism effects that increase the system cost because electricity and hard-to-abate sectors compete for limited renewable resources; the synergy effects that reduce the system cost because electricity and hard-to-abate sectors share hydrogen infrastructure. Yang et al estimated that hydrogen reduces the total investment in hard-to-abate sectors by ~8% in China's near-zero energy system compared to that without hydrogen[11], but do not systematically assess how the interaction between hydrogen infrastructure in the electricity sector (as long-term hydrogen storage and pipeline transportation) and hard-to-abate sectors (as hydrogen fuel and feedstock) influences system costs and investments.

Third, what are the roles of hydrogen pipelines and electric transmission in the zero-carbon energy system? Previous research either assumed zero capital cost and energy loss in both hydrogen pipelines and electric transmission[8,22], or utilized a simplified hydrogen network based on the levelized cost of hydrogen[20]. However, these simplified assumptions overestimated the cost reduction from hydrogen. Given that the underground storage for hydrogen is geographically dependent[27], hydrogen pipelines and electric transmissions are key in balancing hydrogen supply and demand at an hourly temporal resolution between provinces.

Here, we examine the role of green hydrogen in China's zero-carbon electricity system, and its implications for the zero-carbon emission energy system by coupling green hydrogen infrastructure (electrolyzers) for providing long-term storage in the electricity system and fuel and feedstock in hard-to-abate sectors. This study builds upon the GridPath modeling platform by adding the capability to model hydrogen electrolyzers, fuel cells, and combustion turbines, carbon capture capacity with both existing and new coal and gas power plants, direct air capture capacity, and storage and pipeline capacities for both hydrogen and carbon dioxide. Particularly, we modeled the interprovincial transportation of electricity, hydrogen, and carbon dioxide. We selected 3 representative days per month with an hourly resolution to capture diurnal, multi-day, and seasonal variability but limit the burden on computational resources required for co-optimizing both investment and operational costs. With an objective of minimizing total system costs, we simulated the investments and operations of China's electricity system in 2050, subject to various technical, economic, and policy constraints, including a constraint on carbon emissions. We chose 2050 as the target year for the energy system to reach zero carbon emissions because this emission pathway is consistent with China's pledge to reach net zero greenhouse gas emissions by 2060[17]. See the Methods section and Supplementary Information (SI) for more details.

Our baseline scenario is the zero-carbon emission scenario with hydrogen providing long-term storage (ZE scenario, Table 1). We then analyzed four different collections of scenarios designed to explore the trade-offs of different net-zero configurations.

In the first collection, we assessed the impact of carbon constraints (80R, 90R) and the availability of hydrogen (ZE w/o $H_2$) on the cost of the electricity system.

In the second collection, we evaluated how the zero-carbon electricity system is reshaped if one type of electricity and hydrogen infrastructure (technology X) is absent (ZE w/o X scenarios).

In the third collection, to compare the cost-competitiveness of hydrogen with other reliable low-carbon technologies, we examined scenarios where nuclear capacities were expanded (ZE + Nuclear), and CCS and direct air capture (DAC) were installed (ZE + CCS + DAC).

In the fourth collection, to examine the role of hydrogen in the whole energy system, we built a scenario where green hydrogen infrastructure is coupled for providing long-term storage in the electricity system and fuel and feedstock in hard-to-abate sectors (ZE + $H_2$ demand). To assess its advantages, we compared the coupled energy system under (Fig S1), with a decoupled energy system (Fig S2) where green hydrogen infrastructure for hard-to-abate sectors is separate from the green hydrogen infrastructure used for long-term storage in the electricity system (ZE + $H_2$ demand + decouple). To compare the cost-competitiveness of green hydrogen with blue hydrogen (SMR and gasification with CCS), we built a scenario where blue hydrogen is available to meet the fuel/feedstock hydrogen demand (ZE + $H_2$ demand + Blue). We adopted the fuel/feedstock hydrogen demand estimated by the China Hydrogen Allicance[28], which projected hydrogen demand for carbon neutrality considering domestic policy support. Among the projected end-use demand, hydrogen is mainly used in hard-to-abate sectors including industrial heat and feedstock, ammonia and methanol production, and heavy-duty fuel cell electric vehicles, shipping and aviation (Table S1).

**Table 1**. Main scenario descriptions. All scenarios include solar PV, onshore and offshore wind, hydropower, battery storage, and nuclear technologies.

|  | Scenario | 2050 emission | Optional technologies | Hydrogen demand |
| --- | --- | --- | --- | --- |
| Baseline | ZE | Zero | $H_2$ | Electricity sector only |
| Figure 1 | REF<br>80R<br>90R | No carbon target, and 80%, 90% and 100% reduction | $H_2$ | |

| | Scenario | 2050 emission | Optional technologies | Hydrogen demand |
|---|---|---|---|---|
| | REF w/o $H_2$<br>80R w/o $H_2$<br>90R w/o $H_2$<br>ZE w/o $H_2$ | relative to 2020 | w/o H2 | |
| Figure 2 | ZE w/o X | Zero | $H_2$ | |
| Figure 3 | ZE + Nuclear | | 500 GW nuclear[29] | |
| | ZE + CCS +DAC | | $H_2$, coal w/ CCS, natural gas w/ CCS, DAC | |
| Figure 4 | ZE + $H_2$ demand | | $H_2$ (electrolyzers are coupled in electricity and hard-to-abate sectors) | Electricity sector and hard-to-abate sectors |
| | ZE + $H_2$ demand + decouple | | $H_2$ (electrolyzers are separated in electricity and hard-to-abate sectors) | |
| | ZE + $H_2$ demand + Blue | | $H_2$ (electrolyzers are coupled in electricity and hard-to-abate sectors ), SMR w/ CCS, gasification w/CCS, DAC | |

In the ZE scenario, we limited hydropower capacity (including pumped hydro) to 430 GW based on existing and planned capacities[30], and nuclear capacities to 120 GW according to the projection from China's official goal[31]. Following China's policy on coal power plants[32], the maximum coal capacity is limited to 1100 GW. Our analysis includes the following technologies (Table 2): direct carbon capture (DAC), power-to-gas (P2G) and gas-to-power (G2P) technologies, steam methane reforming (SMR), coal gasification, underground storage for hydrogen (i.e., salt caverns) and carbon dioxide (i.e., depleted gas and oil reservoirs, and saline aquifers), and the transportation pipelines for hydrogen and carbon dioxide. Here, DAC is utilized to remove residual emissions left over after CCS deployment. Carbon removal via weathering and reforestation is not considered because they are out of the scope of the energy system. Bioenergy with CCS (BECSS) is not considered in this study because BECCS is expected to incur significant land use change emissions[33,34], and threaten food security[35]. The underground storage capacities for hydrogen[27] and carbon[36] are limited to only those provinces with suitable sites. Due to a lack of data on the provincial potential for China's underground storage, we assume unlimited underground hydrogen and carbon storage capacity in provinces with suitable sites (Table S2). As the geological requirement for hydrogen storage is substantial, we performed sensitivity tests where there is no hydrogen underground storage and where the liquified hydrogen derivatives are available (Fig S11).

Table 2. Technologies assessed in this study to decarbonize China's future electricity system

| Classification | Technology |
| --- | --- |
| Conventional energy | Coal, gas, nuclear, and hydropower |
| Renewable energy | Solar PV, onshore wind, offshore wind |
| Conventional storage | Battery storage, pumped hydro storage |
| Power-to-gas (Power-to-hydrogen) (G2P, green hydrogen) | Electrolyzer |
| Gas-to-power (Hydrogen-to-power) (P2G) | Hydrogen combustion turbine, fuel cell |
| Hydrogen storage | Underground storage (i.e. salt cavern), tank storage |
| Energy and $CO_2$ transport | Electricity grid, hydrogen pipeline, and $CO_2$ pipeline |
| Carbon capture and removal technologies | Carbon capture and storage (CCS), direct air capture (DAC), and underground $CO_2$ storage |
| Fossil-based hydrogen production (gray hydrogen) | Steam methane reforming (SMR), and gasification |

**Effects of hydrogen on deployment and costs**

Hydrogen decreases the levelized cost of a zero-emission electricity system (ZE) by 17% (ranging from 11-20% due to uncertainties in electrolyzers, fuel cells, and storage costs and efficiencies, Fig S3) relative to the system without hydrogen (ZE w/o $H_2$ scenario) (Fig 1f).

A zero-emission electricity system with hydrogen (ZE scenario) decreases the need for wind and solar capacity (4% and 30 %) and yet their generation increases by 6-35% relative to a system without hydrogen (ZE w/o H$_2$ scenario) because long-duration hydrogen storage reduces energy curtailment (Fig 1e). The availability of hydrogen storage also reduces the need for battery storage capacity. Under the ZE scenario, the battery capacity halves relative to the ZE w/o H$_2$ scenario (Fig 1c). Hydrogen displaces battery capacity not only because it balances energy supply and demand across seasons but also between weekdays and weekends. It shifts energy generation from weekends with low electricity demand to weekdays with high electricity demand, providing additional flexibility during weekdays and thus displacing short-duration battery storage (Fig S4). Lastly, hydrogen also avoids substantial electricity grid capacity (40%) in the zero-emission system even though some hydrogen pipeline capacity needs to be built because of heterogeneity in underground storage availability across provinces (Fig 1b). This is because hydrogen storage is more cost-effective than the electricity grid in balancing the mismatch of renewable energy supply and electricity demand.

Without a carbon emissions target, the reference scenario (REF) results in an emissions reduction of 15% in 2050 compared to 2020. Only 48% of energy comes from variable wind and solar resources (Fig 1e) and no hydrogen storage is cost-optimally built (Fig 1d). As the carbon emissions cap is tightened (80R, 90R, and ZE), wind and solar capacity are increasingly installed (Fig 1a) and contribute 80%, 84%, and 88%, respectively, to total energy generation. Commensurately, the deployment of hydrogen storage capacity for long-duration storage services increases exponentially with the 80R, 90R, and ZE scenarios deploying 76 TWh, 152 TWh, and 620 TWh of hydrogen storage capacity, respectively, by 2050 (Fig 1d). That storage capacity in the ZE scenario is equivalent to 18 days of daily average electricity demand. Similarly, hydrogen electrolyzer capacity increases from 600 GW to 950 GW as the carbon cap increases from 80% to 100% (Fig 1e). In all three low-carbon scenarios (i.e., 80R, 90R and ZE), hydrogen decreases system costs. While cost decreases with hydrogen in the 80R and 90R scenarios relative to those without hydrogen are relatively modest (3% and 4%, respectively),

Though hydrogen enables a lower levelized cost under the ZE scenario, reaching a zero-carbon system remains expensive without considering the external costs of climate damages. Compared to the least cost REF scenario, the levelized cost of electricity increases by 72% under the ZE w/o H$_2$ scenario. With hydrogen, the increase in levelized cost relative to the REF scenario is lower but still substantial at 44% (38-53% accounting for uncertainties in hydrogen infrastructure costs).

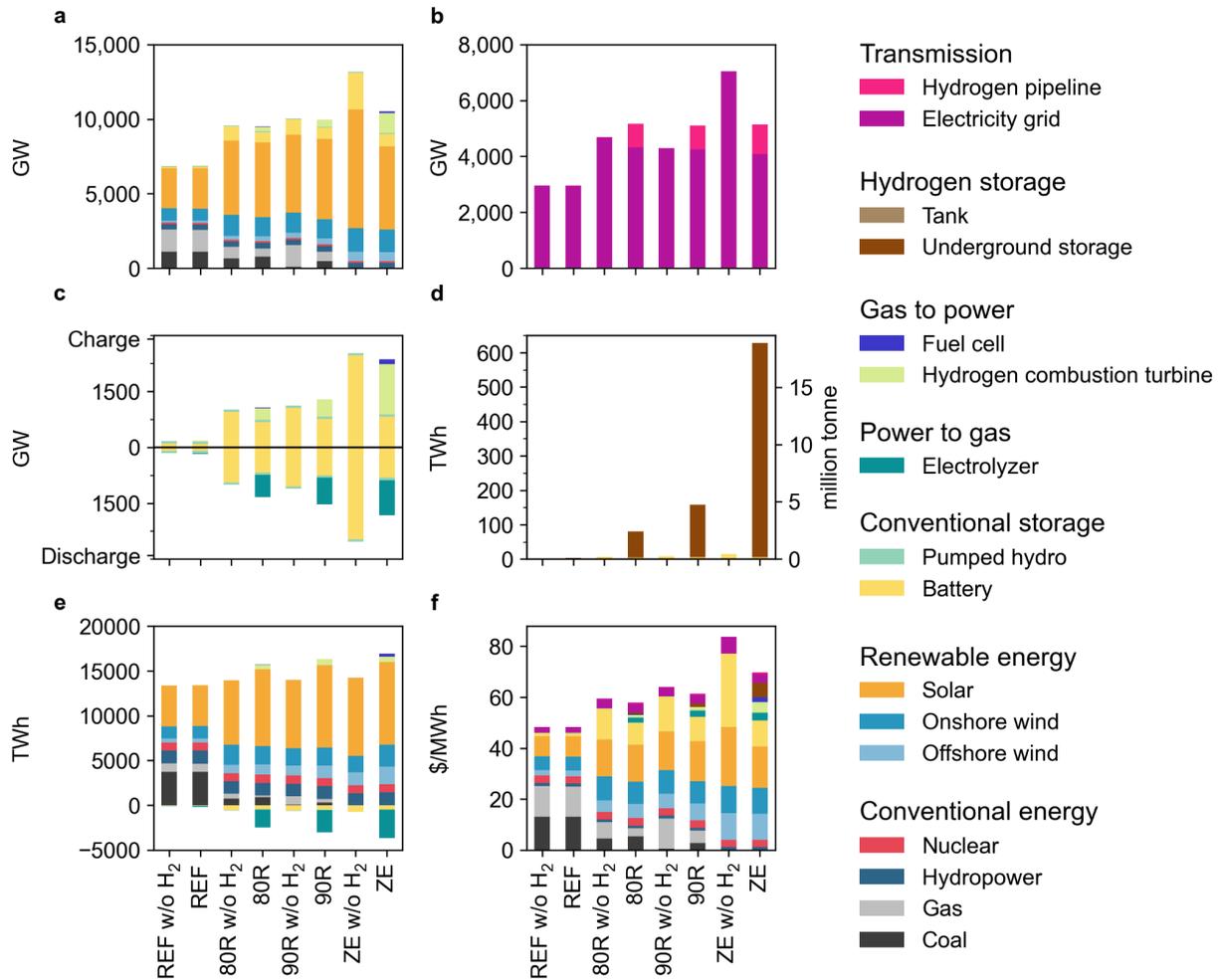

Figure 1. **a**. Generation capacity, **b**. total interprovincial transmission capacity, **c**. power capacity of charging and discharging storage, **d**. energy capacity of storage, **e**. generation, and **f**. levelized cost of electricity under low-carbon scenarios in 2050. In **e**, the negative value for the electrolyzer represents the electricity input for hydrogen. Different colors refer to different technologies. REF refers to the scenario without a carbon emission target. 80R, 90R, and ZE refer to the scenarios where carbon emissions are reduced by 80%, 90%, and 100% relative to 2020 levels.

**Transmission and underground storage are key for the deployment of hydrogen**

The levelized cost of the zero-carbon electricity system depends on both technological and geographical factors. To understand the effect of investments in key hydrogen and electricity infrastructure components in China's zero-emission electricity system, we excluded each component from available investment options and examined their impact on system costs, hydrogen and battery storage requirements, and hydrogen and electricity trade. We found that the expansion of the electricity grid and hydrogen pipelines, and the availability of abundant underground hydrogen storage are key to lowering the levelized cost of electricity (Fig 2).

Not expanding the existing energy transport network (both electric transmission and hydrogen pipelines) increases the levelized cost by 34% compared to the ZE scenario that allows unlimited expansion of both transmission lines and hydrogen pipelines (Fig 2a). The cost increase is mainly driven by the expansion of offshore wind capacities in East Coast China to compensate for the scarce onshore renewable energy resources in those provinces (Fig S5, S6). Additionally, expensive tank storage of hydrogen is deployed in provinces without underground storage sites, which increases the levelized cost of electricity. In provinces with limited high-quality renewable resources, G2P technologies need to have higher hydrogen-to-power efficiencies to meet local demand; therefore, more capacities of high-efficiency but expensive fuel cells are installed (Fig S5, S7). The hydrogen storage capacity increases by 200 TWh (35%) relative to the ZE scenario because the existing limited transmission network cannot balance the variability of renewable energy (Fig 2b).

Solely expanding electricity grids without hydrogen pipelines bumps the levelized cost by only 0.8%, which is similar to the levelized cost when both electricity and hydrogen networks are expanded (Fig 2a). Under the ZE scenario, hydrogen is produced in inland provinces with high-quality wind and solar resources, and sent to East Coast provinces for storage via hydrogen pipelines (Fig S8). Without hydrogen pipelines, hydrogen is produced and stored in the East Coast provinces by importing electricity from inland provinces. In this scenario, the reliance on electricity trade increases. The capacity of the electricity grid increases to 4,800 GW compared to the 4,100 GW in the ZE scenario, and the annual trade of electricity reaches ~7,800 TWh, increasing from 5,800 TWh under the ZE scenario (Fig 2c).

Expanding only hydrogen pipelines without expanding electricity transmission lines increases the levelized cost by 7% (Fig 2a) compared to the ZE scenario. Without transmission lines, more expensive but more efficient fuel cells are built to make the best use of limited high-quality renewables. The energy storage capacity slightly increases to 700 TWh, but the annual trade of hydrogen between provinces doubles to 5,300 TWh compared to that in the ZE scenario (Fig 2c).

Among the hydrogen technologies, inexpensive underground storage (i.e., salt caverns) is the most crucial to lowering the levelized cost (Fig 2a). Without underground storage, no hydrogen technologies are selected because the energy capacity cost of the alternative – tank storage – is greater than the underground storage by 15 times. In the sensitivity test, we found that liquifying all hydrogen into ammonia could increase the levelized cost of electricity by 10% relative to the ZE scenario assuming unlimited hydrogen storage, but decrease the cost by 8% relative to the scenario where underground hydrogen storage is absent (Fig S11). As for the G2P (gas-to-power) technologies, without hydrogen combustion turbines, electricity generation from hydrogen is limited to using only fuel cells, which results in a 7% increase in the levelized cost relative to the ZE scenario. The use of only fuel cells also shrinks the capacity for underground hydrogen storage by a quarter to 430 TWh due to its higher gas-to-power efficiency (Fig 2b).

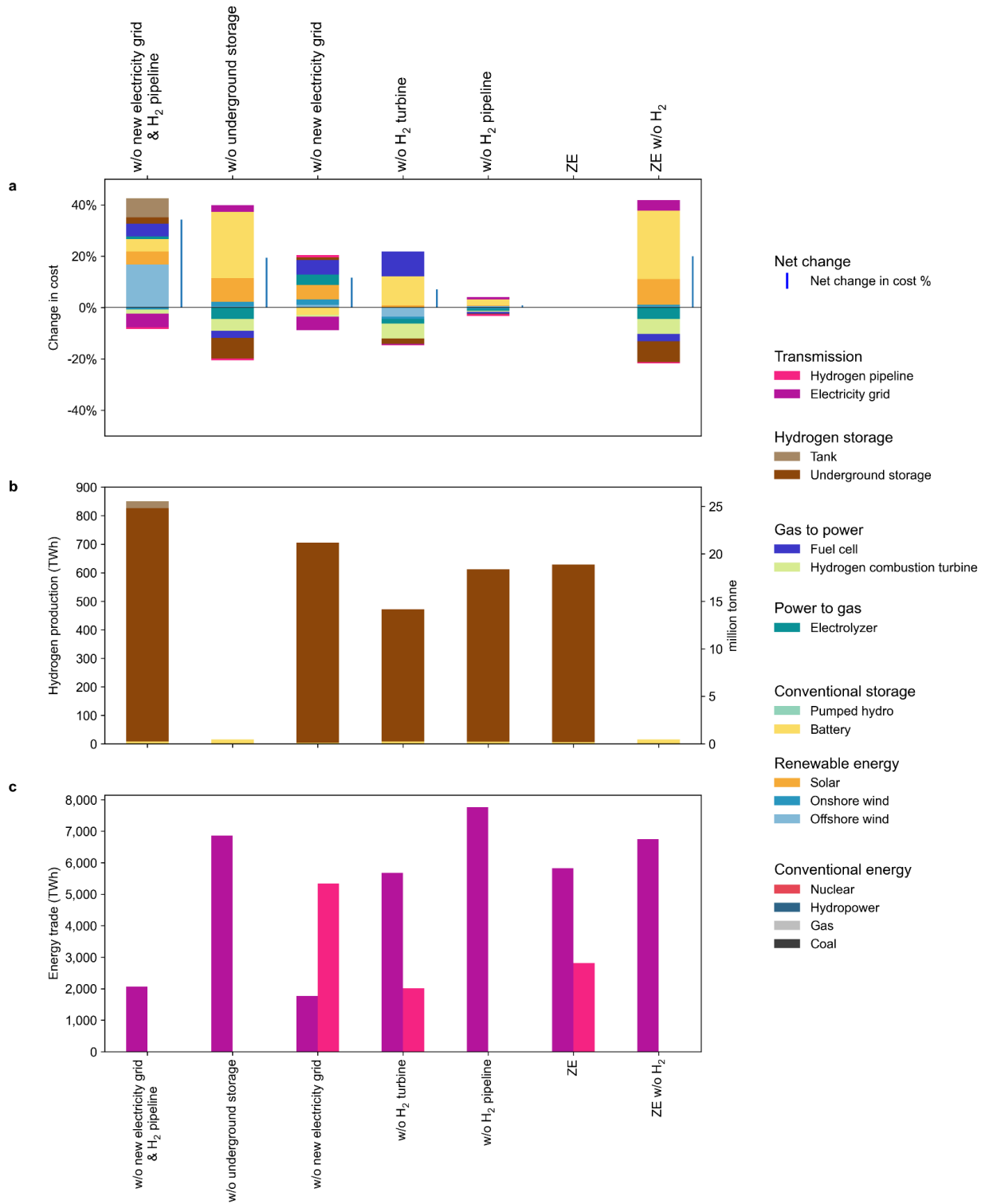

Figure 2. Sensitivity to new investments in electricity and hydrogen infrastructure, **a**. change in cost per unit electricity demand relative to the zero-emission scenario (ZE), **b**. energy capacity of battery and hydrogen storage, and **c**. annual interprovincial trade of electricity and hydrogen. Vertical blue lines in **a** represent the net percentage change in levelized cost.

**Interaction of hydrogen with other low-carbon technologies**

Green hydrogen infrastructure (P2G, storage and G2P) can provide reliable and flexible low-carbon capacity to an electricity system but other technologies including flexible nuclear power plants, and CCS and DAC coupled with coal and gas power plants could also provide similar services. We assessed the impact of these low-carbon technologies on hydrogen infrastructure and other investments and system costs. Under the ZE + Nuclear scenario, we increase the upper limit of available nuclear capacity from 120 GW to 500 GW, and assume that the nuclear fleet is more flexible by allowing it to operate at 50% of its rated capacity instead of 100% under the ZE scenario. For coal and gas power plants coupled with CCS and DAC technologies (ZE + CCS + DAC scenario), we assume an 85% carbon emissions capture rate[37] for CCS and constant costs for CCS and DAC from 2020-2050 (Table S3). Sensitivity tests with a higher carbon capture rate for CCS, and lower capital cost for CCS and DAC are performed in Fig S9.

Overall, adding nuclear, CCS, and DAC technologies reduces the cost of the zero-carbon electricity system. A larger flexible nuclear fleet (ZE + Nuclear scenario) decreases the levelized cost of electricity by 11% compared to the ZE scenario. Allowing investments in CCS and DAC capacities (ZE + CCS + DAC scenario) decreases the levelized cost of electricity by 4% relative to the ZE scenario (Fig 3f). If the capture efficiency of CCS improves from 85% to 95% and capital costs of both CCS and DAC technologies decline by over 50%, the levelized cost decreases by 13% compared to the ZE scenario (Fig S9). Expanding nuclear and installing CCS and DAC at the same time reduces system costs by 15% relative to the ZE scenario, and reduces the capacity requirement of underground hydrogen storage by 70% to 160 TWh (Fig S9).

Installing CCS and DAC capacities increases the electricity generation of coal and natural gas to 5% of the total electricity demand (Fig 3e), generated from a fleet of 1,100 GW compared to none in the ZE scenario (Fig 1a). The cost increase from fossil fuel power plants, CCS and DAC is balanced by lower investment in renewable energy and storage. Relative to the ZE scenario, the solar, wind and battery storage capacities drop by 4-11% (Fig 3a), and the energy capacity of hydrogen storage shrinks by 70% (Fig 3d).

Carbon capture and removal technologies clash with hydrogen. Due to the high operating and capital costs of capturing $CO_2$, fossil power plants are operated not as baseload supply but as marginal capacity operated during peak load hours to minimize the system-wide operation and capital costs. As a result, the need for energy storage during peak hours reduces. Thus, the capacity of electrolyzers decreases by 21% from 950 GW to 740 GW (Fig 3c), and the capacity of hydrogen pipelines decreases by 20% from 1,100 GW to 840 GW (Fig 3b).

Expanding nuclear power plants reduces the capacity of renewable energy. The share of nuclear power in total electricity generation increases from 7% under the ZE scenario to 23% under the ZE + Nuclear scenario. Solar, wind, and battery capacities decrease by 19-39% relative to the ZE scenario. Due to low operating costs, nuclear power plants offer baseload generation, but do not balance the seasonal variation of renewable energy. Bulk hydrogen storage capacities are still required to balance the variation in renewable energy generation across seasons. Relative to the ZE scenario, the energy capacity of hydrogen storage decreases by only 7%.

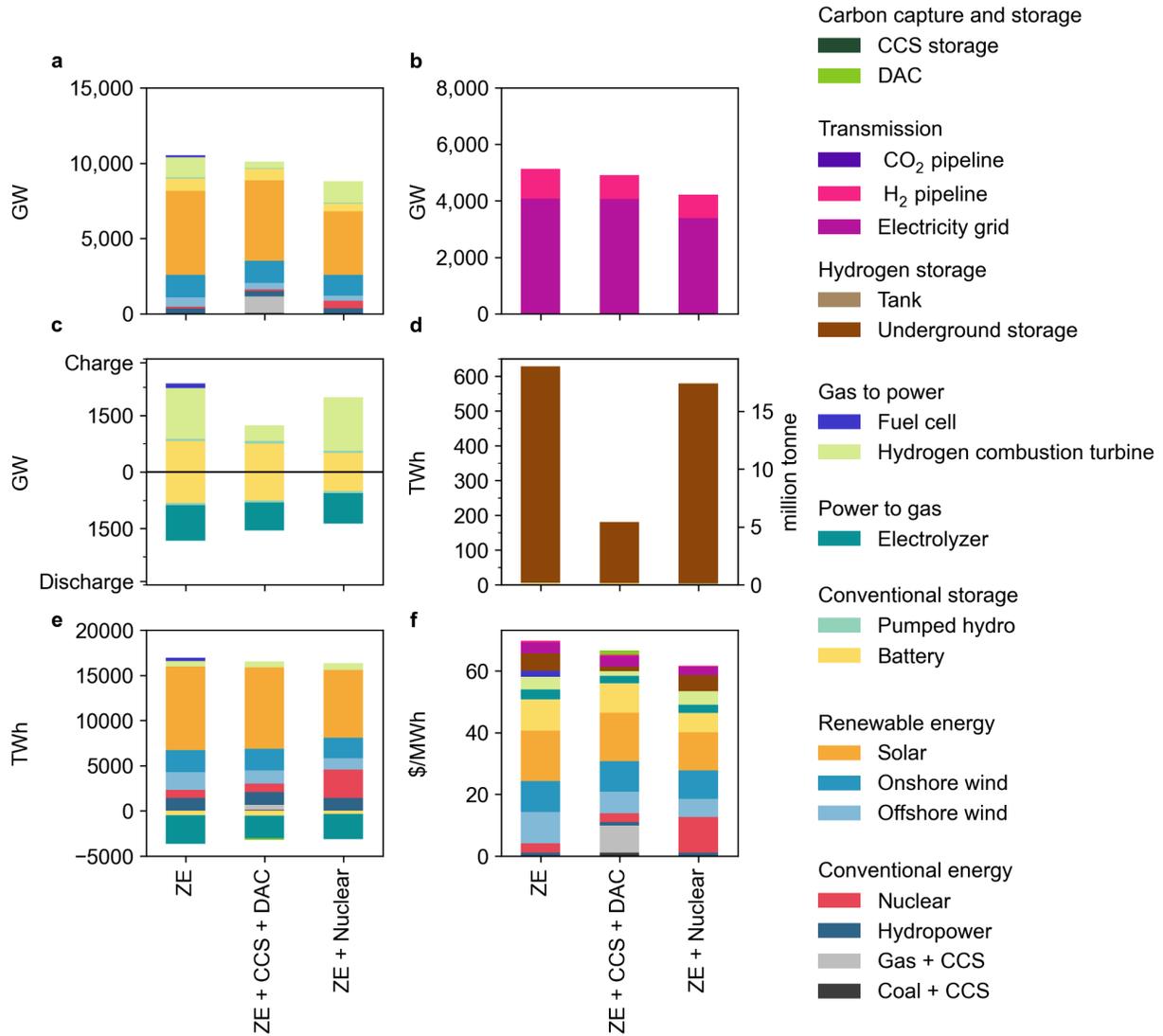

Figure 3. **a**. Generation capacity, **b**. interprovincial transmission capacity, **c**. power capacity of charging and discharging storage, **d**. energy capacity of storage, **e**. annual energy generation, and **f**. levelized cost of electricity under ZE, ZE + CCS +DAC, and ZE + Nuclear scenarios.

**Cost-competitiveness of coupling the electricity and hard-to-abate sectors**

According to China Hydrogen Alliance[28], fuel/feedstock hydrogen demand in China's hard-to-abate sectors (heavy industry, and transportation) and other end-users is expected to rise to 4,100 TWh (124 million tonnes) in scenarios compatible with a net zero energy system (see Method). Among the estimates by different research, we chose this hydrogen demand as the baseline assumption because the number was estimated by China's domestic experts considering the technology-specific policy.

When hydrogen solely serves as long-term storage (ZE scenario), hydrogen infrastructure balances energy supply and demand across seasons caused by the seasonal variation of renewable energy (Fig 4a). Hydrogen is produced and stored between March and August, and discharged in other months. In total, 2,100 TWh of hydrogen is produced, and 980 TWh of electricity is converted by G2P infrastructure.

In contrast, when green hydrogen infrastructure is coupled to provide both long-term storage in the electricity and fuel/feedstock in hard-to-abate sectors (ZE + $H_2$ demand scenario), hydrogen use for long-term storage in the electricity sector decreases to less than 1,000 TWh (Fig 4c), and electricity generated by G2P technologies decreases to 390 TWh (Fig 4b). This is because to meet the additional hydrogen demand from hard-to-abate sectors, more wind and solar capacity is deployed. This additional renewable energy capacity contributes more towards meeting demand in peak hours, and thus reduces the need for hydrogen storage compared to the ZE scenario (Fig 4b). At the same time, more electrolyzers (P2G) are built to meet hydrogen demand from the hard-to-abate sectors, but the capacity of G2P technologies decreases because of the expansion in renewable energy capacities. More importantly, despite the increase in renewable energy capacities, generation curtailment decreases because hydrogen is produced and stored during times of overgeneration of electricity, and can be delivered to hard-to-abate sectors. Relative to the ZE scenario, hydrogen consumed by G2P technologies (hydrogen used for long-term storage) drops by half through coupling the electricity and hard-to-abate sectors (Fig 4d).

Coupling green hydrogen infrastructure in the electricity and hard-to-abate sectors substantially decreases the cost of energy (hydrogen + electricity) relative to scenarios where green hydrogen infrastructures in hard-to-abate and electricity sectors are decoupled. When the electricity system provides hydrogen to the hard-to-abate sectors (ZE + $H_2$ demand scenario), the levelized cost of energy decreases by 7% relative to the case where hydrogen in the electricity system does not serve as fuel/feedstock hydrogen in hard-to-abate sectors (ZE + $H_2$ demand + decouple scenario) (Fig 4c). In the decoupled system, more expensive tank storage are deployed in hard-to-abate sectors to balance the supply and demand of fuel/feedstock hydrogen (Fig 4d).

Introducing green hydrogen from the electricity system into the hard-to-abate sectors remains cost-effective even after allowing blue hydrogen (SMR or coal gasification with CCS and DAC). In the coupled energy system, when hard-to-abate sectors source both green and blue hydrogen (Hydrogen + demand + Blue), the levelized cost of energy is reduced by only 0.2% relative to the ZE + $H_2$ demand scenario. Under the Hydrogen + demand + Blue scenario, blue hydrogen production reaches 280 TWh, and is lower than green hydrogen by an order of magnitude. However, if the electricity system and the hard-to-abate sectors are decoupled, introducing both blue and green hydrogen (Hydrogen + demand + Blue + decouple) reduces the cost by 4% relative to the system with only green hydrogen (Hydrogen + demand + decouple, Fig S10).

In the year 2050, in the coupled energy system, the cost of green hydrogen for hard-to-abate sectors can be similar to and even lower than the cost of coal and gas-based hydrogen (gray hydrogen) without CCS and DAC. In this study, we estimated the lowest cost of gray hydrogen assuming that the capacity utilization rate of SMR and gasification reaches 100%. A range of costs for gray hydrogen is projected given the fossil fuel prices seen during 2018-2023 and the investment cost in 2050 (red lines in Fig 4e and 4f). From a system perspective, we assume that the additional system cost incurred by green hydrogen for fuel/feedstock is the difference in cost between the ZE and the ZE + Hydrogen demand scenarios. This method yields a unit cost of hydrogen of 58 (56-72) $/MWh or 1.93 (1.88-2.40) $/kg. This range of costs for green hydrogen overlaps significantly with the range of costs expected from gray hydrogen (Figs. 4e and 4f). If China's natural gas price rises above ~11 $/mmBtu or coal prices exceed ~2.5 $/mmBtu in 2050, green hydrogen from the zero-emission electricity system is likely to be cheaper than natural gas and coal-based hydrogen (Fig 4e and 4f), a plausible future given the rising real energy prices of fossil fuels.

Because of a lack of data on the potential hourly demand curve of hydrogen, we assumed that the provincial hourly profile of hydrogen demand follows the pattern of electricity demand in each province. We make this assumption on the hourly profile of hydrogen because both electricity and hydrogen demand include end-users like industry and transportation. As the industry sectors may follow a constant hourly demand of hydrogen, we assume a constant hourly hydrogen demand at the provincial level (Fig S10), and find that the constant hourly demand assumption does not substantially change the cost of energy.

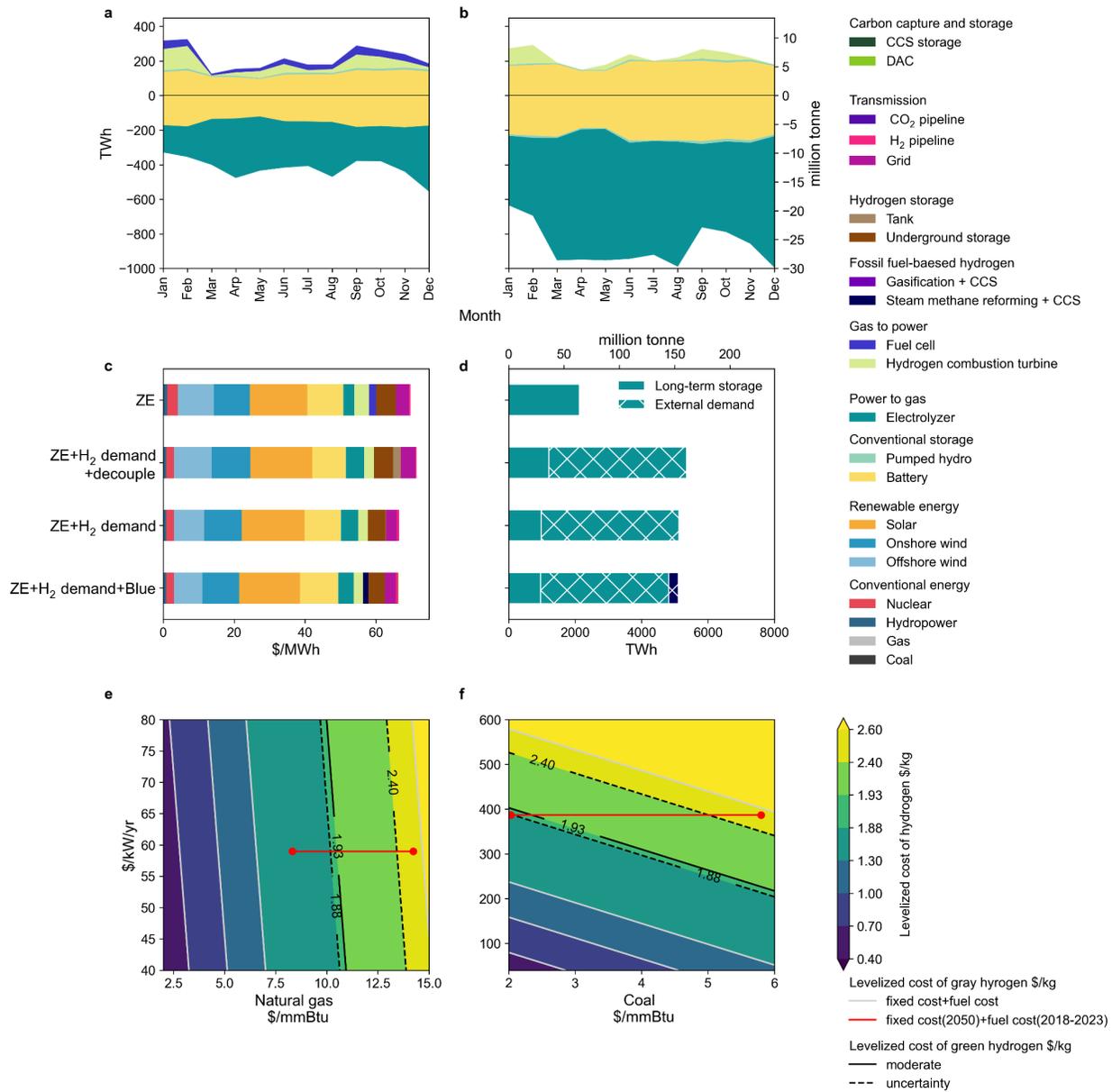

Figure 4. Generation curve of hydrogen by month under **a**. ZE and **b**. ZE + $H_2$ demand scenarios. **c**. Levelized cost of energy (both electricity and hydrogen production for hard-to-abate sectors) under the ZE, ZE + $H_2$ demand + decouple, ZE + $H_2$ demand, and ZE + $H_2$ demand + Blue scenarios. **d**. Hydrogen consumed by long-term storage and fuel/feedstock hydrogen demand. Contour maps showing the levelized cost under different fixed costs and fuel costs for **e**. natural gas-based gray hydrogen (SMR) and **f**. coal-based gray hydrogen (gasification). The gray solid diagonal lines show contours of the levelized cost of gray hydrogen under different fixed costs and fuel costs. The solid black diagonal lines represent the combination of fixed cost and fuel

cost where the levelized cost of gray hydrogen equals the levelized cost of green hydrogen. The diagonal dashed black lines represent the combination where the levelized cost of the gray hydrogen equals the lower higher bound of the levelized cost for green hydrogen. The red horizontal lines represent the levelized cost of gray hydrogen given the projected 2050 fixed cost (387 $kW/year) and 2018-2023 fuel costs[38].

**Discussion**

Hydrogen has increasingly gained a focus on power and energy systems decarbonization. As the world's largest greenhouse gas emitter, China has pledged to achieve net zero greenhouse gas emissions by 2060 to mitigate climate change. This pledge requires that the energy system reaches zero carbon emissions before 2050[17]. However, without technological innovations, achieving a zero-carbon electricity system can be expensive. Here, we find that including hydrogen as long-term storage can reduce the levelized cost of electricity by 17% in a zero-carbon electricity system where fossil fuels are entirely phased out by 2050. We also find that a zero-carbon electricity system can satisfy hydrogen demand in hard-to-abate sectors at a reasonable cost. If the price of natural gas increases to over 13 $/mmBtu and the price of coal increases to over 2.5 $/mmBtu, hydrogen produced by zero-carbon electricity is likely to be cheaper than fossil-based hydrogen. Relative to the cost-optimal scenario with no carbon cap, however, the levelized cost of electricity in the ZE still increases by 44% assuming a moderate rate of cost decline in electrolyzers, fuel cells, and hydrogen storage (Table S4-S7), and potentially imposes a large financial burden for investors and consumers.

Several factors will shape the prospects of hydrogen use in China. First, the availability of underground storage is the limiting factor for using hydrogen as long-term storage. Without underground storage, using tanks to store hydrogen will be a more expensive option. Salt caverns are the best sites to store hydrogen, and saline aquifers are potential candidates[39,40]. Our research estimates that the underground hydrogen storage capacity should reach ~600 TWh. While Europe has found a total of 84.8 PWh hydrogen storage sites,[41] the potential for salt cavern capacity is unknown in China. The latest research shows that the operating salt cavern in China can store only 0.3 TWh hydrogen, which is 3 orders of magnitude lower than the needed capacity.[27] One solution to address the shortage of hydrogen underground storage is by liquifying

H$_2$ into hydrogen derivatives. These hydrogen derivatives have higher energy density and are much cheaper for tank storage compared to hydrogen.

Adding other reliable low-carbon technologies like nuclear power plants to the ZE electricity system further reduces the cost. Carbon capture (CCS) and removal (DAC) technologies replace hydrogen technologies in the electricity system, as they both serve as moderators for the variability of renewable energy. If CCS and DAC technologies are available, the levelized cost of electricity decreases by 4% relative to the zero-carbon electricity system with hydrogen, and the hydrogen storage capacity decreases. This antagonistic effect indicates that the prospect of hydrogen as long-term storage depends on whether the Chinese government decides to fully phase out fossil fuel power plants.

While these pathways with nuclear, CCS, and DAC technologies may be economically feasible, they pose socio-political challenges. Introducing new nuclear power plants raises risk concerns in the public[42]. Coal power plants cause ~100,000 annual premature deaths due to air pollution[43]. Reliance on natural gas casts doubt on energy security, given that 40% of China's natural gas was imported in 2020, while supplying 3% of electricity generation[44]. Other pathways that include hydropower and bioenergy expansion, which were not examined in this study, also face socio-political risks. Expanding hydropower capacities results in the change of ecosystem[45] and the resettlement of communities[46,47]. Moreover, to meet China's carbon-neutral target, bioenergy reduces food supply by 8% if assuming the current level of food import[35].

Expanding the electricity transmission network is crucial for hydrogen use. Cheap electricity will need to be transported to regions with underground hydrogen storage sites. Solely expanding electricity transmission lines can achieve similar cost reduction as expanding both pipelines and the electricity transmission lines. By retrofitting China's existing natural gas pipelines to hydrogen, the cost can be further reduced.

More importantly, an operation system that manages hydrogen production in both electricity and non-electricity systems is necessary. Incorporating the electricity and hard-to-abate sectors under a coupled system reduces the cost by 20%, compared to the case where the two systems work independently. This coupled system replaces the traditional fossil-based hydrogen pathways, facilitating decarbonization in the hard-to-abate sectors. Given China's reliance on expensive imported natural gas, scaling up hydrogen using natural gas is not economically or politically feasible compared to the green hydrogen pathway. We estimated the cost of fuel/feedstock hydrogen to be 1.9 $/kg from a system perspective, which includes the cost reductions in the electricity system due to the coupling of the hydrogen infrastructure (see method). By considering the system cost increase incurred by fuel/feedstock hydrogen, our estimate of hydrogen cost is consistent with other forecasts[48,49]. If we only calculate the cost of the electrolyzer and the average cost of electricity without considering the cost reduction in the electricity system, the average cost of green hydrogen goes over 3 $/kg..

Our study has several limitations. First, the future cost of hydrogen and other renewable energy technologies is highly uncertain. In our deterministic model, we performed sensitivity tests considering the low and high projections of technology cost, as well as the low energy conversion efficiency for hydrogen technologies (Fig. S1). While the cost estimates may be uncertain due to the uncertainties in technology cost and efficiency, the conclusion that hydrogen reduces the levelized cost of the electricity and energy system remains robust.

Second, the investment cost for new end-use technologies in hard-to-abate sectors (e.g., direct reduced iron) is not considered when calculating the levelized cost of energy. This may lead to an underestimate of the total system cost. However, in all our scenarios, we assume the same fuel/feedstock hydrogen demand, which translates to the same investment for end-use technologies across scenarios. Therefore, our conclusion that coupling green hydrogen in the electricity system and hard-to-abate sectors holds robust.

Third, the fuel/feedstock hydrogen demand is uncertain due to various assumptions in a net-zero energy system. To ensure that green hydrogen is capable of meeting high hydrogen demand, we used a relatively high estimate of hydrogen demand from the China Hydrogen Association[28]. Adopting a lower projection of fuel/feedstock hydrogen demand will decrease the levelized cost of energy because of a decreased need for low-quality renewable resources.

Fourth, we assume unlimited capacities of transmission lines can be built between provinces, which results in 3,000-4,000 GW of interprovincial transmission lines. Currently, the average capacity of an ultra-high voltage direct current transmission line (UHVDC) is ~10 GW[24]. In our study, we have about 100 interprovincial transmission corridors. In reality, this means about 3-4 transmission lines need to be built between provinces by 2050. The number of transmission lines can be lower in the real world, as capacity, geophysical and social constraints may limit the expansion of transmission lines. Limiting the expansion of transmission lines increases the importance of hydrogen in a zero-emission system as shown in Figure 2.

Last, the zero-carbon emission system here only refers to the scope 1 emissions of $CO_2$ within the electricity or energy system, and excludes upstream or fugitive emissions of methane, $H_2$, etc, embedded $CO_2$ emissions of PV and wind turbine manufacturing, the carbon sink from land use (weathering, reforestation, etc). Mitigation cost estimates for a zero-carbon emissions system increase when upstream emissions are considered[50]. Future research may consider the fugitive emissions of hydrogen and methane to better estimate the cost for a net-zero energy system, and their risks in offsetting the mitigation efforts of carbon emissions.[51] While our study focuses on the relative change of the cost, the exclusion of upstream emission does not change our conclusion.

**Method**

**Electricity model**

We used the GridPath model, an open-source power system model, to optimize the total investment and operation costs of electricity infrastructure in China in 2050.[52,53] In our GridPath-China model, the 31 provinces in China are classified into 32 load zones, where Inner Mongolia is split into an Eastern Inner Mongolia load zone and a Western Inner Mongolia load zone. The energy transport between load zones is also optimized by our power system model.

Each month in the model has three representative days with hourly resolution. To capture diurnal and seasonal dynamics while reducing the computational burden and maintaining problem tractability, we used 36 representative days in our capacity expansion model considering the modeling solving time, which may miss some extreme days with wind or solar drought. Using the 36 representative days requires 2-48 hours to solve the model, considering the complexity of energy transmission (grid, hydrogen pipelines and carbon pipelines) and technology portfolios. The representative days are selected based on maximum, median, and minimum electricity demand in each month. The representative days with minimum electricity demand represent weekend days, while the maximum and median days represent weekdays. In a 100% decarbonized electricity system, Bistline found that using 24 representative days (one peak day and one average day) has a decent performance in simulating the investment choices in renewable energy and carbon removal technologies[54]. To ensure reliability during peak load hours, we assumed a planning reserve margin of 15% of the peak load. To examine the reliability of the zero-carbon electricity system, we ran an operation model with fixed capacities in 8760 hours (Table S8). We found that in a zero-carbon electricity system with hydrogen, the demand can be met every hour.

Total coal capacity is constrained to less than 1100 GW in all scenarios based on the National Development and Reform Commission's policy to avoid over-capacity of coal generation.[32] The minimum generation level assumed is 100% of rated capacity for nuclear power plants, 40% for coal power plants and gas turbines, and 45% for combined cycle gas turbines. The hourly ramp rate of the rated capacity is 30% for coal power plants and 60% for gas power plants.[55] The energy loss of transmission lines and hydrogen pipelines is shown in Table S6.

Existing generation capacities for all technologies following the scheduled retirement in 2050 (Table S9), and monthly average capacity factors of hydropower were compiled from He et al. (2016)[56]. Except for hydropower, other existing power plants retire in 2050. We collected the existing and planned hydropower and pumped hydro capacities from Global Energy Monitor.[30] We collected data for existing transmission lines from State Grid[57] and Southern Grid[58]. Hourly demand, hourly generation capacity factors and resource potential for solar, onshore and offshore wind were collected from Abhyankar et al.[59] The distribution of salt caverns and carbon storage sites were collected from Zhu et al[27] and Fan et al[36].

In this study, we improved the GridPath model by adding a hydrogen module, including P2G, G2P and hydrogen storage technologies. We also added hydrogen pipelines into the model, which allows the transportation of hydrogen across load zones. Another improvement is the addition of a CCS/DAC module, which optimizes the investment and operation of carbon capture. The CCS/DAC module considers the storage of carbon emissions, and the transport of carbon emissions through pipelines. Detailed formulations are presented in Supporting Information and Kamal et al[52].

**Cost assumptions**

China-specific renewable energy capital costs were collected from the International Renewable Energy Agency[60], and China-specific capital costs of battery and pumped hydro storage technologies, and fossil-fuel technologies were collected from Zhuo et al (Table S4)[24]. The O&M costs for renewables, storage and fossil fuels were collected from Zhuo et al[24]. We then applied normalized cost projection curves from 2020 to 2050, derived from the moderate cost projection from the NREL 2023 Annual Technology Baseline (ATB) database[61] to the China-specific technology costs. In this study, we used the 2020 dollar value and an 8% discount rate.

As hydrogen and CCS have not been developed at a large scale in China, the costs for the electrolyzers, salt cavern, hydrogen storage tank, CCS underground storage (depleted oil and gas reservoirs and saline aquifers), hydrogen and CCS pipeline, and DAC were from the Danish Energy Agency (Table S3-S6)[62]. The costs for CCS were collected from the Global CCS Institute (Table S3)[63]. The projected cost for transmission lines was collected from Grid Project Construction Cost Analysis in the 12th Five-year Period (Table S6)[64]. The fuel costs of coal, natural gas, and uranium were curated by He et al (2016, 2020)[55,56] and Luo et al[65] (Table S7).

**Scenario**

The base scenario, "ZE", in this analysis is China's 2050 electricity system with zero carbon emissions. In this scenario, no fossil fuel generation is allowed. The hydropower and pumped hydro capacities (430 GW) include existing and planned capacities. The nuclear capacity is capped at 120 GW. The 'ZE w/o $H_2$' scenario refers to an electricity system with zero carbon emissions, but without any hydrogen generation or infrastructures.

We include various scenarios to assess the uncertainty associated with input costs and technologies. These include zero-emission scenarios with low (ZE + Low cost) and high-cost ('ZE + High cost') hydrogen technologies. The 'ZE + High cost + Low efficiency' assesses a zero-emission scenario with high-cost and low-efficiency hydrogen technologies.

We then assess the impact of constraints on the various types of electricity and hydrogen infrastructure on costs and emissions compared to the base zero-emission (ZE) scenario. These zero-emission scenarios include those without new electricity transmission lines ('ZE w/o new electricity grid'), without new H2 pipelines ('ZE w/o $H_2$ pipeline'), without both new electricity transmission lines and H2 pipelines ('ZE w/o new electricity grid and $H_2$ pipeline'), without underground storage for hydrogen ('ZE w/o underground storage'), and without new hydrogen combustion turbines ('ZE w/o $H_2$ turbine').

The reference ('REF') scenario has no constraint on $CO_2$ emissions from China's power system. In '90R' and '80R' scenarios, $CO_2$ emissions from the electricity system are reduced by 90% and 80% respectively relative to 2020.

To understand the effect of other low-carbon technologies on hydrogen infrastructure investments and system costs, we include scenarios with both CCS combined with DAC and nuclear. We simulate a zero-emission scenario with CCS and DAC technologies ('ZE + CCS + DAC') By assuming that the cost of CCS and DAC stay the same as 2020 and a CCS capture rate of 85%. To assess the effects of technology and cost improvements, we also include a low-cost, high-efficiency CCS and DAC scenario ('ZE + CCS + DAC'), where the costs of CCS and DAC decline by 50% and the CCS capture rate improves to 95%.

In the zero-emission scenario with expanded nuclear power plant deployment ('ZE + Nuclear'), we allow nuclear capacity to expand from 120 GW in the base scenario to 500 GW. We assume a minimum generation level for nuclear power plants of 50% instead of 100% under the base ZE scenario. To assess the effect of higher costs of nuclear energy ('ZE + Nuclear (high cost)' scenario), we assume the relatively conservative nuclear cost projection from NREL while keeping the same input assumptions as the 'ZE + Nuclear' scenario. We also include a zero-emission scenario with no constraints on nuclear capacity expansion ('ZE + Unlimited nuclear' scenario). Lastly, we simulate a zero-emission scenario with both CCS combined with DAC and a high nuclear scenario ('ZE + CCS + DAC + Nuclear') where the new nuclear buildout follows the assumptions under the 'ZE + Nuclear' scenario, and CCS and DAC infrastructure follows the assumption under the 'ZE + CCS +DAC' scenario.

Finally, we include a zero-emission scenario where China's 2050 electricity system serves additional hydrogen demand from hard-to-abate sectors including industry (e.g., direct reduced iron, methanol, ammonia, and industry heat), transportation (heavy-duty vehicle, shipping and aviation), and building ('ZE + Hydrogen demand'). To examine the cost implications of this coupled system, we develop a case where the electricity system does not directly serve hydrogen demand in hard-to-abate sectors ('ZE + Hydrogen demand + decouple'). In this 'ZE + Hydrogen demand + decouple' scenario, hard-to-abate sectors meet their hydrogen by building electrolyzers that don't serve long-term storage in the electricity sector. To compare the green hydrogen scenarios with fossil-based hydrogen (Blue hydrogen), we build two additional zero-emission scenarios where the additional hydrogen demand from hard-to-abate sectors can also be sourced from SMR and/or coal gasification, and the fossil fuel emissions are removed by CCS and DAC. The electricity and hard-to-abate sectors in these two scenarios are either coupled ('ZE + Hydrogen demand + Blue') or decoupled ('ZE + Hydrogen demand + Blue

+ decouple'). We assume an 85% CCS capture rate and that costs of CCS and DAC remain unchanged from 2020 levels.

**Electricity and hydrogen demand**

The 2050 provincial-level electricity demand for China was collected from Abhyankar et al, which projected higher electricity demand (15 PWh) due to electrification compared to other projections.[59] The projected electricity demand doubles the demand in 2020 (~7.6 PWh). In our main scenarios ('ZE'), when hydrogen is only used as long-term storage in the electricity system, fuel/feedstock hydrogen demand from hard-to-abate sectors is assumed to be zero.

In the 'ZE + Hydrogen demand' scenario, the national hydrogen demand (excluding the hydrogen demand in the electricity system) in a net-zero energy system is equal to the demand estimated by the China Hydrogen Association (124 million tonnes, Table S1), which is reported by the International Energy Agency.[28] This estimate considers the domestic policy support for hydrogen technologies, and thus is higher than the estimate from the IEA. The lower heating value of hydrogen (120 MJ/kg) is used to convert hydrogen between mass and energy values. The hourly curve of hydrogen demand follows the normalized hourly curve of electricity demand, and the spatial distribution of hydrogen demand follows that of electricity. A sensitivity test with a flat demand curve for hydrogen demand is shown in Fig S10.

The levelized cost of hydrogen (LCOH) includes the capital cost of electrolyzers (Cap_E) and the cost of electricity. The electricity consumed by hydrogen in the electricity system (ED) and hard-to-abate sectors (HD) is calculated by considering the power-to-hydrogen efficiency ($\eta$) and the levelized cost of electricity (LCOE).

$$LCOH = \frac{Cap\_E + LCOE \times (ED+HD)/\eta}{ED+HD} \quad (1)$$

The electricity consumed by hydrogen in hard-to-abate sectors (HD) is calculated by considering capital cost of electrolyzers in hard-to-abate sectors ($Capacity\ cost_{HD}$), and the power-to-hydrogen efficiency ($\eta$) and the levelized cost of electricity (LCOE).

$$LCOH = \frac{Capacity\ cost_{HD} + LCOE \times HD/\eta}{HD} \quad (2)$$

In the energy system, if the electricity system and hard-to-abate sectors are coupled, the same electrolyzer capacity produces hydrogen for both long-term storage and fuel/feedstock. The average cost of hydrogen equation (1) is equal to the average cost of fuel/feedstock hydrogen in hard-to-abate sectors derived by equation (2).

However, equation (2) didn't consider the cost reduction in the electricity system incurred by coupling green hydrogen in the electricity system and hard-to-abate sector. Compared to the grey hydrogen, coupling green hydrogen in the electricity system and hard-to-abate sectors reduces the cost of the electricity system.

To measure the unit cost of fuel/feedstock hydrogen, we used the additional system incurred by fuel/feedstock hydrogen to measure the unit cost of fuel/feedstock hydrogen ($LCOH_{fuel/feedstcok}$) in the coupled energy system. $HTA^{ZE + Hydrogen\ demand}$ is the capital cost of fuel/feedstock hydrogen in the hard-to-abate sectors. $Ele^{ZE}$ and $Ele^{ZE + Hydrogen\ demand}$ is the cost of the electricity system under the ZE and ZE + Hydrogen demand scenario. The $Ele^{ZE + Hydrogen\ demand} - Ele^{ZE}$ is the electricity cost for the fuel/feedstock hydrogen. The LCOH for fuel/feedstock hydrogen can thus be calculated as follows:

$$Total\ cost^{ZE + Hydrogen\ demand} = Ele^{ZE + Hydrogen\ demand} + HTA^{ZE + Hydrogen\ demand} \quad (3)$$

$$Total\ cost^{ZE} = Ele^{ZE} \quad (4)$$

$$LCOH = \frac{HTA^{ZE + Hydrogen\ demand} + (Ele^{ZE + Hydrogen\ demand} - Ele^{ZE})}{HD} \quad (5)$$

$$LCOH = \frac{Total\ cost^{ZE + Hydrogen\ demand} - Total\ cost^{ZE}}{HD} \quad (6)$$

where $Total\ cost^{ZE + Hydrogen\ demand}$ is the total cost to meet both electricity and hydrogen demand under the ZE + Hydrogen demand scenario, and $Total\ cost^{ZE}$ is the total cost to meet only electricity demand under the ZE scenario. In Figure 4, when calculating the levelized cost of gray hydrogen, the capacity factor of SMR and gasification is assumed to be 100%.

As our model only includes the capital cost of electrolyzer/SMR/gasification in hard-to-abate sectors, the capacity cost of hydrogen ($Capacity\ cost_{HD}$) is equal to the capital cost of the hard-to-abate sector ($HTA^{ZE + Hydrogen\ demand}$).

$$Capacity\ cost_{HD} = HTA^{ZE + Hydrogen\ demand} \quad (7)$$

In the energy system, the total electricity costs ($Ele^{ZE + Hydrogen\ demand}$) consist of two parts: the electricity cost for producing hydrogen in hard-to-abate sectors ($Ele\_HD^{ZE+Hydrogen\ demand}$), and the electricity cost for meeting electricity demand ($Ele\_Ele^{ZE+Hydrogen\ demand}$).

$$Ele^{ZE + Hydrogen\ demand} = Ele\_HD^{ZE+Hydrogen\ demand} + Ele\_Ele^{ZE+Hydrogen\ demand} \quad (8)$$

$$Ele\_HD^{ZE+Hydrogen\ demand} = LCOE \times HD/\eta \quad (9)$$

When the hard-to-abate sectors are independent of the electricity sector, the cost increase in the energy system equals the capacity and electricity cost of hydrogen. The LCOH in equation (6) is equivalent to LCOH in equation (2).

When the hard-to-abate sectors are independent of the electricity sector, the cost in the electricity system under the ZE + Hydrogen demand scenario ($Ele\_Ele^{ZE+Hydrogen\ demand}$) is the same as that under the ZE scenario $Ele^{ZE}$.

$$Ele\_Ele^{ZE+Hydrogen\ demand} = Ele^{ZE} \quad (10)$$

Then the electricity cost used to produce the fuel/feedstock hydrogen is equal to:

$$LCOE \times HD/\eta = Ele^{ZE + Hydrogen\ demand} - Ele\_Ele^{ZE+Hydrogen\ demand} \quad (11)$$

$$LCOE \times HD/\eta = Ele^{ZE + Hydrogen\ demand} - Ele^{ZE} \quad (12)$$

Thus, the equation (2) and (6) are equivalent when the hard-to-abate sectors and the electricity sectors are independent of each other.

$$LCOH = \frac{Total\ cost^{ZE + Hydrogen\ demand} - Total\ cost^{ZE}}{HD} = \frac{Capacity\ cost_{HD} + LCOE \times HD/\eta}{HD} \quad (13)$$

When the hard-to-abate sectors are coupled with the electricity sector, LCOH in equation (6) measures the capacity and electricity cost of hydrogen, and the cost reduction in the electricity system.

Due to the coupling of the electricity system with hard-to-abate sectors, the electricity cost under the $ZE + Hydrogen\ demand$ scenario will be lower than the electricity cost under the ZE scenario.

$$Ele\_Ele^{ZE+Hydrogen\ demand} < Ele^{ZE} \quad (14)$$

Therefore, the electricity cost for hydrogen in equation (4) will be lower than the electricity cost for hydrogen in equation (8).

$$Ele^{ZE + Hydrogen\ demand} - Ele\_Ele^{ZE} > Ele^{ZE + Hydrogen\ demand} - Ele^{ZE} \quad (15)$$

$$LCOE \times HD/\eta > Ele^{ZE + Hydrogen\ demand} - Ele^{ZE} \quad (16)$$

By considering the cost reduction to the electricity system, the LCOH in equation (6) is lower than the conventional LCOH in equation (2).

$$\frac{Capacity\ cost_{HD} + LCOE \times HD/\eta}{HD} > \frac{Total\ cost^{ZE + Hydrogen\ demand} - Total\ cost^{ZE}}{HD} \quad (17)$$

**Data and code availability**

The code and data supporting this article are available at 10.5281/zenodo.13937324

**Declaration of Interest**

The authors declare no competing interests.

**Supplementary Information**

Document S1. Figures S1–S11, Tables S1-S9 and Note S1.

# Supplementary Information

Haozhe Yang[1,2,*], Gang He[3], Eric Masanet[1,4], Ranjit Deshmukh[1,5,*]

[1] Bren School of Environmental Science and Management, University of California Santa Barbara, Santa Barbara, California, 93106, USA

[2] Andlinger Center for Energy and Environment, Princeton University, Princeton, New Jersey, 08540, USA

[3] Marxe School of Public and International Affairs, Baruch College, City University of New York, New York, 10010, USA

[4] Department of Mechanical Engineering, University of California Santa Barbara, CA, 93106, USA

[5] Environmental Studies Program, University of California Santa Barbara, Santa Barbara, California, 93106, USA

Correspondence: haozheyang@princeton.edu, rdeshmukh@ucsb.edu

**Note S1. Mathematical equations for the hydrogen and CCS modules.**

**Mathematical equations**
**Index**
- **Temporal**
  In the model, the index for each time point is t, the index for each time horizon is h, and the index for each period is p.

  hours_in_tmp is how many hours a timepoint represents.

  tmp_weight is the weight of each time point, equal to 8760/(number of time points × hours_in_tmp).

- **Load zone**
  z

- **Project**
  prj

- **Hydrogen module**

**Electricity to hydrogen (e.g., electrolyzer)**

Per load zone z, at each time point t, the electricity input (Fuel_ele) is less than the rated capacity of P2G (C_ele_H2) in time period p.

$$\text{Fuel\_ele}(t,z) \leq \text{C\_ele\_H2}(p,z)$$

Per load zone z, at each time point t, the conversion from power to hydrogen (Gen_ele_H2) is equal to the electricity input multiplied by the electricity-to-hydrogen efficiency (p2g_$\eta$)

$$\text{Gen\_ele\_H2}(t,z) = \text{Fuel\_ele}(t,z) \times \text{p2g\_}\eta(t,z)$$

**CH$_4$ or coal to hydrogen (SMR)**

Per load zone z, at each time point t, the production of H$_2$ from SMR or gasification (Gen_Fuel_H2) is less than the installed capacity of SMR or gasification at each period p (Cap_Fuel_H2).

$$\text{Gen\_Fuel\_H2}(t,z) \leq \text{Cap\_Fuel\_H2}(p,z)$$

**Hydrogen storage**

Within a time horizon, per load zone z, at time point t-1, after discharging (Gen_H2_Discharge) from and charging (Gen_H2_Charge) to the H$_2$ storage project, we derive the H$_2$ storage (Storage_H2) at time point t. The parameters–efficiency of charging and discharging–are represented as stor_charging_$\eta$ and stor_discharging_$\eta$.

$$\begin{aligned}
\text{Storage\_H2}(t,z) = \ & \text{Storage\_H2}(t-1,z) + \\
& \text{Gen\_H2\_Charge}(t-1,z) * \text{stor\_charging\_}\eta * \text{hours\_in\_tmp} * \text{tmp\_weight} - \\
& \text{Gen\_H2\_Discharge}(t-1,z) / \text{stor\_discharging\_}\eta * \text{hours\_in\_tmp} * \text{tmp\_weight}
\end{aligned}$$

Per load zone z, at each time point t, the H$_2$ storage should be lower than the rated H$_2$ storage capacity (H2_capacity_mwh) in time period p.

$$\text{Storage\_H2}(t,z) \leq \text{H2\_capacity\_mwh}(p,z)$$

**Hydrogen to electricity (e.g., fuel cell)**

Per load zone z, at each time point t, the generation of electricity from hydrogen (Gen_H2_ele) is the hydrogen input (Fuel_H2) multiplied by the parameter, the efficiency of G2P (g2p_$\eta$).

$$\text{Gen\_H2\_ele}(t,z) = \text{Fuel\_H2}(t,z) * g2p\_\eta$$

Per load zone z, at each time point t, the generation of hydrogen-to-power is less than the rated capacity of G2P in period p.

$$\text{Gen\_H2\_ele}(t,z) \leq \text{C\_H2\_ele}(p,z)$$

**H2 pipeline**

H2 pipeline flow from load zone z_s to load zone z_d (H2_flow) is lower than the rated capacity of pipelines at period p (H2_pipeline_max). H2_flow can be negative or positive.

$$\text{H2\_flow}(t,z\_s,z\_d) \leq \text{H2\_pipeline\_max}(p,z\_s,z\_d)$$
$$\text{H2\_flow}(t,z\_s,z\_d) \geq -\text{H2\_pipeline\_max}(p,z\_s,z\_d)$$

H2_pipeline_min(a,b)

Losses for the 'from' load zone of hydrogen (z_s) through the pipeline (H2_loss, which is a non-negative variable), assuming a loss rate of H2_loss_factor

$$\text{H2\_from\_loss}(t,z\_s,z\_d) \geq -\text{H2\_flow}(t,z\_s,z\_d) * \text{H2\_loss\_factor}(z\_s,z\_d)$$
$$\text{H2\_from\_loss}(t,z\_s,z\_d) \leq \text{H2\_pipeline\_max}(z\_s,z\_d) * \text{H2\_loss\_factor}(z\_s,z\_d)$$

Similarly, losses for the 'to' load zone of hydrogen (z_d). Here, the fugitive emission of $H_2$ is not considered in the carbon accounting of the system.

$$\text{H2\_to\_loss}(t,z\_s,z\_d) \geq \text{H2\_flow}(t,z\_s,z\_d) * \text{H2\_loss\_factor}(z\_s,z\_d)$$
$$\text{H2\_to\_loss}(t,z\_s,z\_d) \leq \text{H2\_pipeline\_max}(z\_s,z\_d) * \text{H2\_loss\_factor}(z\_s,z\_d)$$

At time point t, H2_export is the total export of $H_2$ from load zone z_s to other load zones.

$$Export\_H2(t, z\_s) = \sum_{z \in z\_d} [H\_flow(t, z\_s, z) + H2\_from\_loss(t, z\_s, z)]$$

At time point t, H2_import is the total import of $H_2$ to load zone z_d from other load zones.

$$Import\_H2(t, z\_d) = \sum_{z \in z\_s} [H\_flow(t, z, z\_d) - H2\_to\_loss(t, z, z\_d)]$$

- **CCS and DAC module**

Per load zone z, at each time point t, for projects that install CCS (prj_ccs), $CO_2$ captured by CCS (CCS_Capture_tonne) is less than the rated capacity (CCS_Capture_capacity_tonne)

CCS_Capture_tonne(prj_css,t,z) ≤ CCS_Capture_capacity_tonne(prj_ccs, p,z)

Per load zone z, at each time point t, $CO_2$ capture by DAC (DAC_Capture_capacity_tonne) is less than the rated capacity (DAC_Capture_capacity_tonne)

DAC_Capture_tonne(t,z) ≤ DAC_Capture_capacity_tonne(p,z)

Per load zone z, at time point t, for project prj_ccs, $CO_2$ captured by CCS is lower than the carbon emissions multiplied by the parameter carbon removal rate (carbon_removal_$\eta$ , %). Carbon emissions from prj_ccs equals the fuel burned (Fuel_burn) multiplied by the emission factor of the fuel (EF_fuel).

CCS_Capture_tonne(prj_ccs,t,z) <= carbon_removal_$\eta$ * Fuel_burn (prj_ccs,t,z) * EF_fuel

Total carbon storage over all the time points is less than the aggregate of carbon storage capacity **(Carbon_Storage_Capacity_Tonne)**

$$\sum_{t} [\sum_{prj\_ccs} CCS\_Capture\_tonne(prj\_ccs, t, z) + DAC\_Capture\_tonne(t, z)]$$

$$\leq Carbon\_Storage\_Capacity\_Tonne\ (z)$$

Per time point t, $CO_2$ flow from load zone z_s to load zone z_d (CO2_flow) is less than the rated capacity of $CO_2$ pipelines (CO2_pipeline_max) at period p. CO_flow can be positive or negative.

CO2_flow (t,z_s,z_d) <= CO2_pipeline_max(p,z_s,z_d)
CO2_flow (t,z_s,z_d) >= -CO2_pipeline_max(p,z_s,z_d)

Per time point t, CO$_2$ losses for the 'from' load zone z_s (CO$_2$_loss, which is a non-negative value), assuming a leakage rate of CO2_loss_factor (default value is 0).

CO2_from_loss (t,z_s,z_d) >= - CO2_flow(t,z_s,z_d) * CO2_loss_factor(z_s,z_d)
CO2_from_loss (t,z_s,z_d) <= CO2_pipeline_max(p,z_s,z_d)* CO2_loss_factor(z_s,z_d)

Per time point t, CO$_2$ losses for the 'to' load zone z_d.

CO2_to_loss (t,z_s,z_d) >= CO2_flow(t,z_s,z_d) * CO2_loss_factor(t,z_s,z_d)
CO2_to_loss(t,z_s,z_d) <= CO2_pipeline_max(p,z_s,z_d)*CO2_loss_factor(t,z_s,z_d)

CO2_export is the total CO$_2$ exported to all other load zones z from load zone z_s

$$\text{Export\_CO2}(t, z\_s) = \sum_{z \in z\_d} [CO2\_flow(t, z\_s, s) + CO2\_from\_loss(t, z\_s, z)]$$

CO2_import is the total CO$_2$ imported from all other load zones z to load zone z_d

$$\text{Import\_CO2}(t, z\_d) = \sum_{z \in z\_s} [CO2\_flow(t, z, z\_d) - CO2\_to\_loss(t, z, z\_d)]$$

Carbon leakage is the summation of CO$_2$ leakages from pipelines. In this study, we assume that the carbon leakage from pipelines is zero.

$$\text{Carbon\_leakage} = \sum_{z, z\_s, z\_d} [CO2\_to\_loss(t, z\_s, z\_d) + CO2\_from\_loss(t, z\_s, z\_d)]$$

- **Carbon emissions**

Per project prj in load zone z, at each time point t, carbon emission (Carbon_emisson) is equal to the carbon emissions from the fuel burned (Fuel burn).

Carbon_emission (prj,t,z) =Fuel burn (prj,t,z) * EF_fuel

Specifically,
per project prj_ccs in load zone z, at each time point t, carbon emission (Carbon_emisson) is equal to the carbon emissions from the fuel burned minus the carbon emissions captured by CCS.

$$\text{Carbon\_emission}(prj\_ccs, t, z) = \text{Fuel burn}(prj\_ccs, t, z) * EF\_fuel - \text{CCS\_Capture\_tonne}(prj\_ccs, t, z)$$

Per project prj_dac in load zone z, at each time point t, carbon emission (Carbon_emisson) is equal to the negative of carbon emissions captured by DAC.

$$\text{Carbon\_emission}(prj\_dac, t, z) = - \text{DAC\_Capture\_tonne}(prj\_dac, t, z)$$

- **Net power output by project**

Generally, per project (prj) in load zone z, at each time point t, the net power output (Net_power) is the gross power output.

$$\text{Net\_power}(prj, t, z) = \text{Gross\_power}(prj, t, z)$$

Specifically, for P2G projects prj_p2g, the net power is,

$$\text{Net\_power}(prj\_p2g, t, z) = -\text{Ele\_input}(t, z)$$

For G2P projects prj_g2p, the net power output is,

$$\text{Net\_power}(prj\_p2g, t, z) = \text{Gen\_H2\_ele}(t, z)$$

For electricity storage projects prj_stor, the net power output is the net of discharge (Discharge) and charge (Charge).

$$\text{Net\_power}(prj\_p2g, t, z) = \text{Discharge}(t, z) - \text{Charge}(t, z)$$

Per project prj_ccs in load zone z, at each time point t, the net power output is the gross power output (Gross_power) less electricity consumed by CCS. ele_per_tonne_CCS is electricity consumption to capture one tonne $CO_2$ by CCS.

Net_power (prj_ccs,t,z) = Gross_power(prj_css,t,z) - CCS_Capture_Tonne(prj_ccs,t) * ele_per_tonne_CCS

Per project prj_dac in load zone z, at each time point t, the net power output is the negative of electricity consumed by DAC. ele_per_tonne_DAC is electricity consumption to capture one tonne $CO_2$ by DAC.

Net_power (p_dac,t,z) = - DAC_Capture_Tonne(t,z) * ele_per_tonne_DAC

- **Electricity system constraint**

**Carbon constraint**
Per load zone z, the total of end-of-pipe emissions and carbon leakage should be lower than the carbon cap (carbon_cap) in period p.

$$\sum_{prj,t} Carbon\_emission(prj, t, z) + Carbon\_leakage\ (z) <= carbon\_cap(p,z)$$

**Carbon supply-demand balance**
Per load zone z, at time point t, the aggregate of carbon captured and imported should all be equal to the carbon stored and exported.

$$\sum_{prj\_ccs} CCS\_Capture\_tonne(prj\_ccs, t, z) + CCS\_DAC\_tonne(t, z) + Import\_CO2(t, z)$$
$$= CCS\_Storage\_tonne(t, z) + Export\_CO2(t, z)$$

**Hydrogen supply-demand balance**
Per load_zone z, hydrogen generation is equal to hydrogen consumption. Load_H2(t,z) is the hydrogen demand from non-electricity sectors at time point t.

| | |
|---|---|
| Load_H2(t,z) | Non-electricity hydrogen demand |
| +Export_H2(t,z)= | hydrogen export |
| +Gen_Fuel_H2(t,z) | hydrogen from gas or coal |

| +Gen_ele_H2(t,z) | hydrogen generation |
| +Gen_H2_Discharge(t,z)-Gen_H2_Charge(p,z) | hydrogen loss from storage |
| -Fuel_H2 (t,z) | hydrogen used for electricity |
| +Import_H2(t,z) | hydrogen import |

**Power supply-demand balance**

Load(t,z)                                                                                        electricity demand
+ Export_load(t,z)=                                                                    electricity export

$\sum_{p} Net\_power(p, t, z)$                                        electricity from all projects

+Import_load(t)                                                                              electricity import

- **Objective function: minimize total cost**

Total cost = investment cost + operation costs

Investment costs from new projects = Aggregate annualized investment costs from new projects

Operation costs from new projects
= O&M operation costs from electricity generation
+ O&M operation costs from $H_2$ generation
+ O&M operation costs from $CO_2$ capture
+ fuel costs from electricity generation
+ fuel costs from $H_2$ generation

**Table S1. Description of the fuel/feedstock hydrogen demand by sector[1]**

| Sector | Demand (million tonne) | Description |
| --- | --- | --- |
| Industry | 77 | Industrial heating and feedstocks; refining; ammonia and methanol production |
| Transportation | 41 | Buses and trucks (fuel cell), shipping (fuel cell and synthetic fuel), and aviation |

| | | | | | |
|---|---|---|---|---|---|
| | | | | (synthetic fuel) | |
| Building | | 6 | | Blending in the natural gas network; power and heat co-generation | |

**Table S2. Sites for underground hydrogen and carbon storage**

| Technology | Province | Reference |
|---|---|---|
| Underground hydrogen | Anhui, Hebei, Henan, Shandong, Hubei, Hunan, Sichuan, Jiangsu, Jiangxi, Guangdong, Yunnan, Chongqing, Gansu, Inner Mongolia | Zhu et al[2] |
| Underground $CO_2$ | Onshore: Anhui, Chongqing, Jiangsu, Jilin, Hebei, Heilongjiang, Hubei, Henan, Inner Mongolia, Liaoning, Qinghai, Sichuan, Tianjin, Xinjiang<br><br>Offshore: Fujian, Guangxi, Guangdong, Hainan, Jiangsu, Liaoning, Shandong | Fan et al[3] |

**Table S3. Cost and parameter assumptions for CCS and DAC technologies.**

| $ million per tonne per hour | | 2020 | | 2050 | | Electricity (MWh/ tonne) | Lifetime (year) | Ref |
|---|---|---|---|---|---|---|---|---|
| | | Capital | O&M | Capital | O&M | | | |
| CCS | Coal | 3.1 | 0.02 | 1.4 | 0.008 | 0.15 | 40 | Global CCS Institute[4] |
| | Natural gas | 3 | 0.03 | 1.4 | 0.01 | 0.16 | 40 | |
| | SMR/ gasification | 0.8 | 0.0003 | 0.4 | 0.0003 | 0.61 | 40 | |

| CCS storage | Onshore | 2.8 | 0.0 | 2.3 | 0.0 | | 30 | Danish Energy Agency[5] |
|---|---|---|---|---|---|---|---|---|
| | Offshore | 4.7 | 0.0 | 3.8 | 0.0 | | | |
| DAC | | 7 | 0.4 | 2 | 0.4 | 1.5 | 30 | |

Table S4. Cost assumptions for conventional electricity generation and storage technologies. The original data in RMB is converted to the US dollar by assuming that the exchange rate between RMB and the US dollar is 6.5:1. O&M costs are all from NREL ATB databases[6].

| $/kW or $/kWh | | 2020 | | 2050 | | Lifetime (years) | Ref |
|---|---|---|---|---|---|---|---|
| | | Capital | O&M | Capital | O&M | | |
| Coal | Coal | 622 | 10 | 495 | 8 | 40 | Zhuo[7] |
| | IGCC | 1090 | 19 | 1030 | 18 | | |
| Natural gas | CT | 367 | 15 | 286 | 13 | | |
| | CCGT | 409 | 20 | 318 | 15 | | |
| Nuclear | | 2462 | 86 | 1739 (Conservative: 2297) | 86 | 60 | |

| | | | | | | |
|---|---|---|---|---|---|---|
| Hydropower | | 2240 | 40 | No new hydropower | 100 | |
| PV | Utility | 628 | 10 | 308 | 6 | IRENA[8] |
| | Residential | 746 | 10 | 292 | 4 | 20 |
| Wind | Onshore | 1157 | 22 | 784 | 17 | 30 |
| | Offshore | 2857 | 67 | 1947 | 44 | |
| Battery | Power ($/kW) | 492 | 25 | 412 | 20 | 15 | Zhuo[7] |
| | Energy ($/kWh) | 246 | | 111 | | |
| Pumped hydro (10-hour) | | 918 | 23 | No new pumped hydro | 100 | |

**Table S5. Cost and parameter assumptions for hydrogen technologies. The original data in Euro is converted to the US dollar by assuming that the exchange rate between the Euro and the US dollar is 1.1:1.**

| $/kW or $/kWh | | Capital | | | O&M | Efficiency | Lifetime (year) | Ref |
|---|---|---|---|---|---|---|---|---|
| | | 2020 | 2050 | | | | | |
| | | | Low | Medium | High | | | | |
| Electrolyzer | | 1245 | 283 | 311 | 385 | 4% | 70% Low:55% | 25 | Danish Energy Agency[5] |
| Hydrogen combustion turbine | | 315 | 305 | | | Gas CT | 40% | 40 | McPherson[9] |
| Fuel cell | | 1562 | 600 | 962 | 1087 | 5% | 60% Low: 40% | 15 | Danish Energy Agency[5] |
| Hydrogen storage ($/kWh) | Tank | 69 | 25 | 25 | 42 | 2% | 90% | 25 | |
| | Underground | 3.6 | 1.2 | 1.6 | 2.2 | 0 | 99% | 100 | |

**Table S6. Cost and parameter assumptions for transmission and pipelines. The 2020 exchange rate between the Euro and the US dollar is 1.1:1.**

|  | $/MW/km $/tonne/km | Loss/1000 km | Lifetime (year) | Ref |
|---|---|---|---|---|
| Transmission line $/MW/km | 340 | 5.3% | 20 | Electric Power Planning Design General Institute[10] |
| Hydrogen pipeline $/MW/km | 226 | 1.3% | 50 | Danish Energy Agency[5] |
| $CO_2$ pipeline $/tonne/hour/km | 14,000 | 0 |  |  |

**Table S7. Fuel cost. The data for the fuel cost is from He et al.[11] and Luo et al[12].**

| Fuel | Region | $/mmBtu |
|---|---|---|
| Coal | Anhui | 4.61 |
| Coal | Beijing | 4.93 |
| Coal | Chongqing | 3.82 |
| Coal | Inner Mongolia | 2.80 |
| Coal | Fujian | 5.77 |
| Coal | Gansu | 3.60 |
| Coal | Guangdong | 6.17 |
| Coal | Guangxi | 6.17 |

| Coal | Guizhou | | 6.17 |
|---|---|---|---|
| Coal | Hainan | | 6.17 |
| Coal | Hebei | | 4.93 |
| Coal | Heilongjiang | | 4.23 |
| Coal | Henan | | 4.87 |
| Coal | Hubei | | 4.41 |
| Coal | Hunan | | 4.72 |
| Coal | Jiangsu | | 5.77 |
| Coal | Jiangxi | | 4.72 |
| Coal | Jilin | | 3.15 |
| Coal | Liaoning | | 3.38 |
| Coal | Ningxia | | 3.19 |
| Coal | Qinghai | | 3.09 |
| Coal | Shaanxi | | 3.15 |
| Coal | Shandong | | 4.93 |
| Coal | Shanghai | | 5.77 |
| Coal | Shanxi | | 4.22 |
| Coal | Sichuan | | 3.82 |
| Coal | Tianjin | | 4.93 |
| Coal | Xinjiang | | 2.60 |
| Coal | Yunnan | | 6.17 |
| Coal | Zhejiang | | 5.77 |

| Gas | National | 13.68 |
|---|---|---|
| Uranium | National | 0.82 |

**Table S8. Examine the demand curtailment of the capacity expansion model**

| Scenario | Temporal resolution | Demand curtailment |
|---|---|---|
| 'ZE' | 36 days[1] | 0% |
|  | 8760[2] | 0% |

[1] Optimized the capacity investment by using 36 days in a whole year,

[2] Simulated the operation with fixed capacities across 8760 hours.

**Table S9. Existing capacities (MW) for all technologies in 2050 according to retirement year.**

|  | Non-pumped hydropower | Pumped hydropower | Nuclear | Coal | Gas | PV | Wind |
|---|---|---|---|---|---|---|---|
| Anhui | 2950 | 3680 | | | | | |
| Beijing | 739 | 822 | | | 0 | | |
| Chongqing | 9438 | 1200 | | | | | |

| East Inner Mongolia | 721 | |
| --- | --- | --- |
| Fujian | 11586 | 1400 |
| Gansu | 7800 | |
| Guangdong | 12379 | 12971 |
| Guangxi | 15481 | |
| Guizhou | 20493 | |
| Hainan | 1395 | 400 |
| Hebei | 1615 | 8739 |
| Heilongjiang | 795 | 1200 |
| Henan | 3796 | 1320 |
| Hubei | 35125 | 1381 |
| Hunan | 14111 | |
| Jiangsu | 2649 | 3542 |
| Jiangxi | 5617 | 1200 |
| Jilin | 4602 | 1700 |
| Liaoning | 2721 | 3000 |
| Ningxia | 333 | |
| Qinghai | 11423 | |
| Shaanxi | 3733 | 1400 |
| Shandong | 1090 | 4000 |
| Shanxi | 2506 | 2485 |
| Sichuan | 95661 | 2 |
| Tianjin | 17 | |
| Tibet | 2663 | 90 |
| West Inner Mongolia | 1731 | 1285 |
| Xinjiang | 6424 | 1200 |
| Yunnan | 77107 | |
| Zhejiang | 10405 | 7323 |

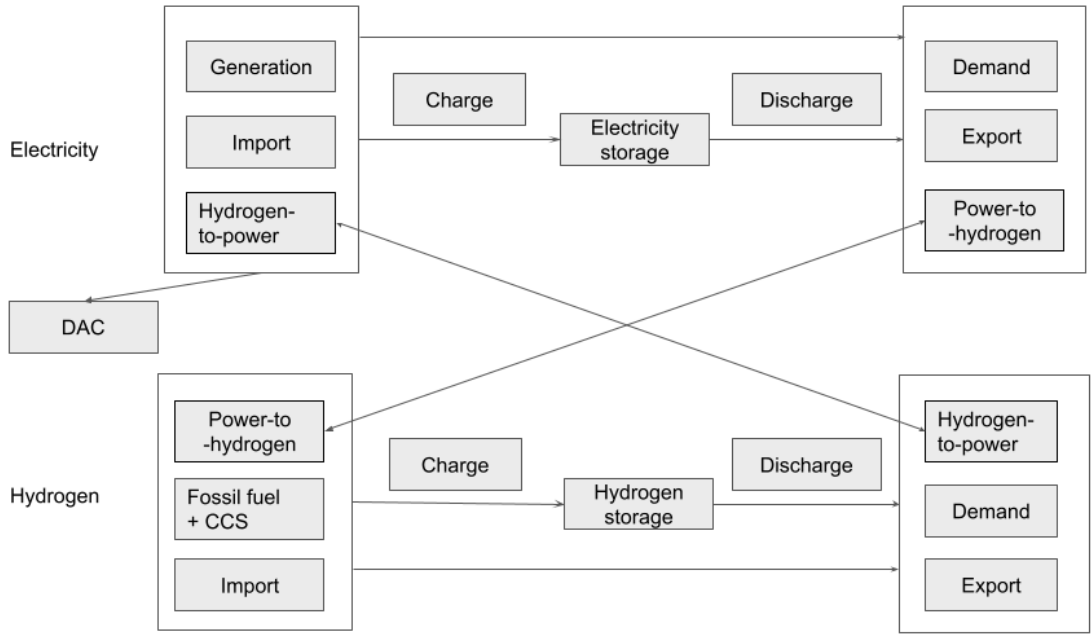

**Figure S1. Flow chart of the coupled electricity system and hard-to-abate sectors.**

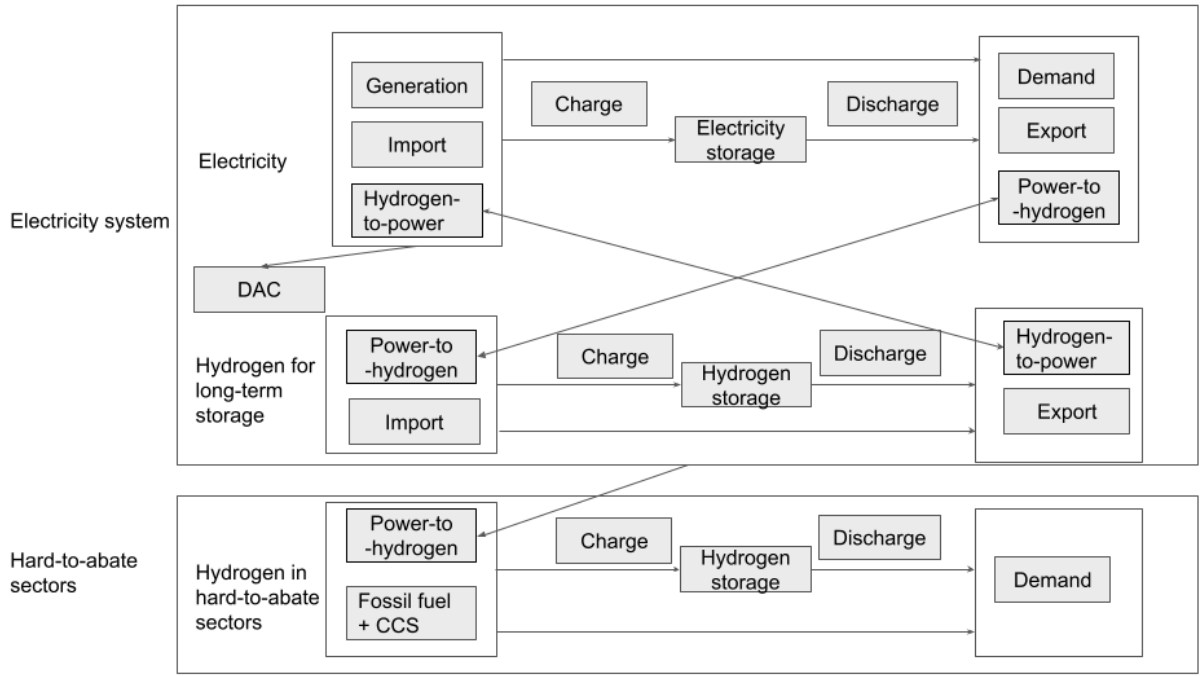

**Figure S2. Flow chart of the decoupled electricity system and hard-to-abate sectors.**

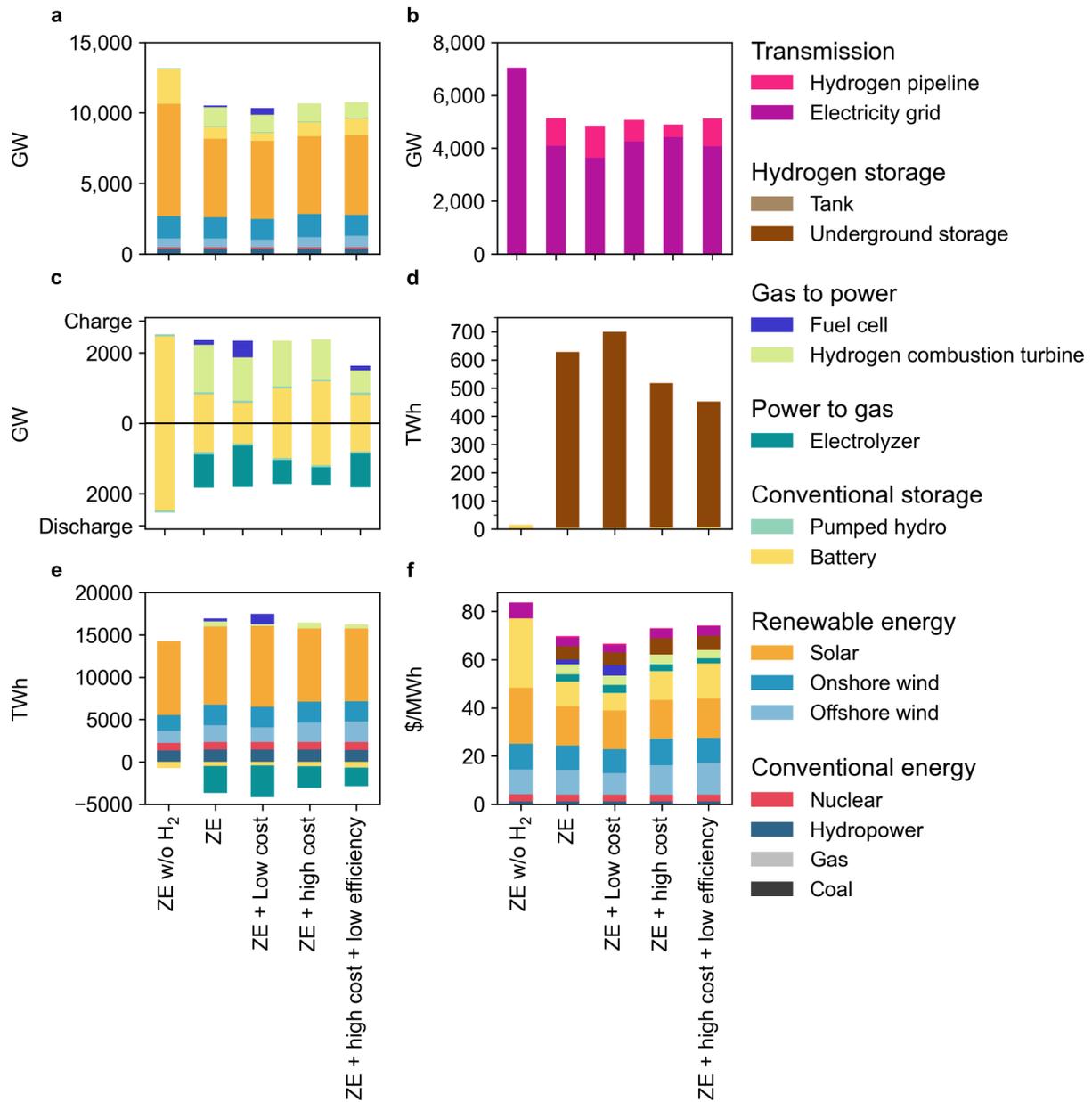

**Figure S3. a**. Generation capacity, **b**. transmission capacity, **c**. power capacity of storage, **d**. energy capacity of storage, **e**. generation, and **f**. levelized cost of electricity under low-carbon scenarios in 2050. Different colors refer to different technologies. ZE + Low cost refers to the ZE scenario with low-cost hydrogen technologies. ZE + High cost refers to the ZE scenario with high-cost hydrogen technologies. ZE + High cost + Low efficiency refers to the ZE scenario with high-cost and low-efficiency hydrogen technologies

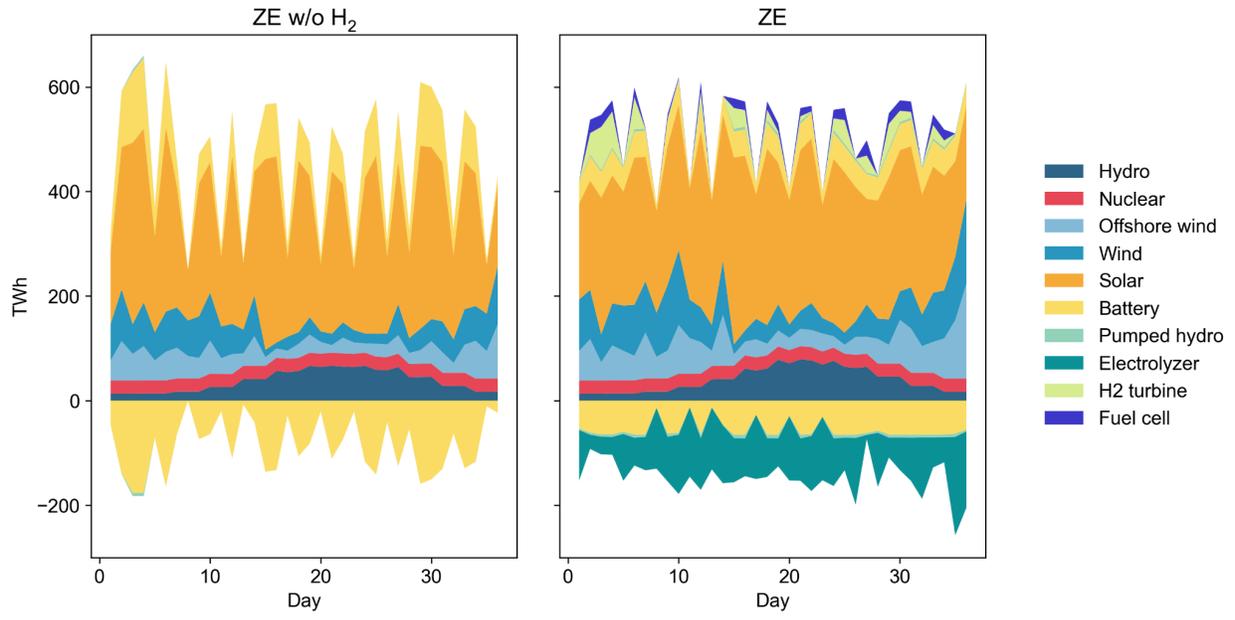

**Figure S4. Daily generation of electricity in the 36 representative days under ZE w/o H$_2$ and ZE scenarios.**

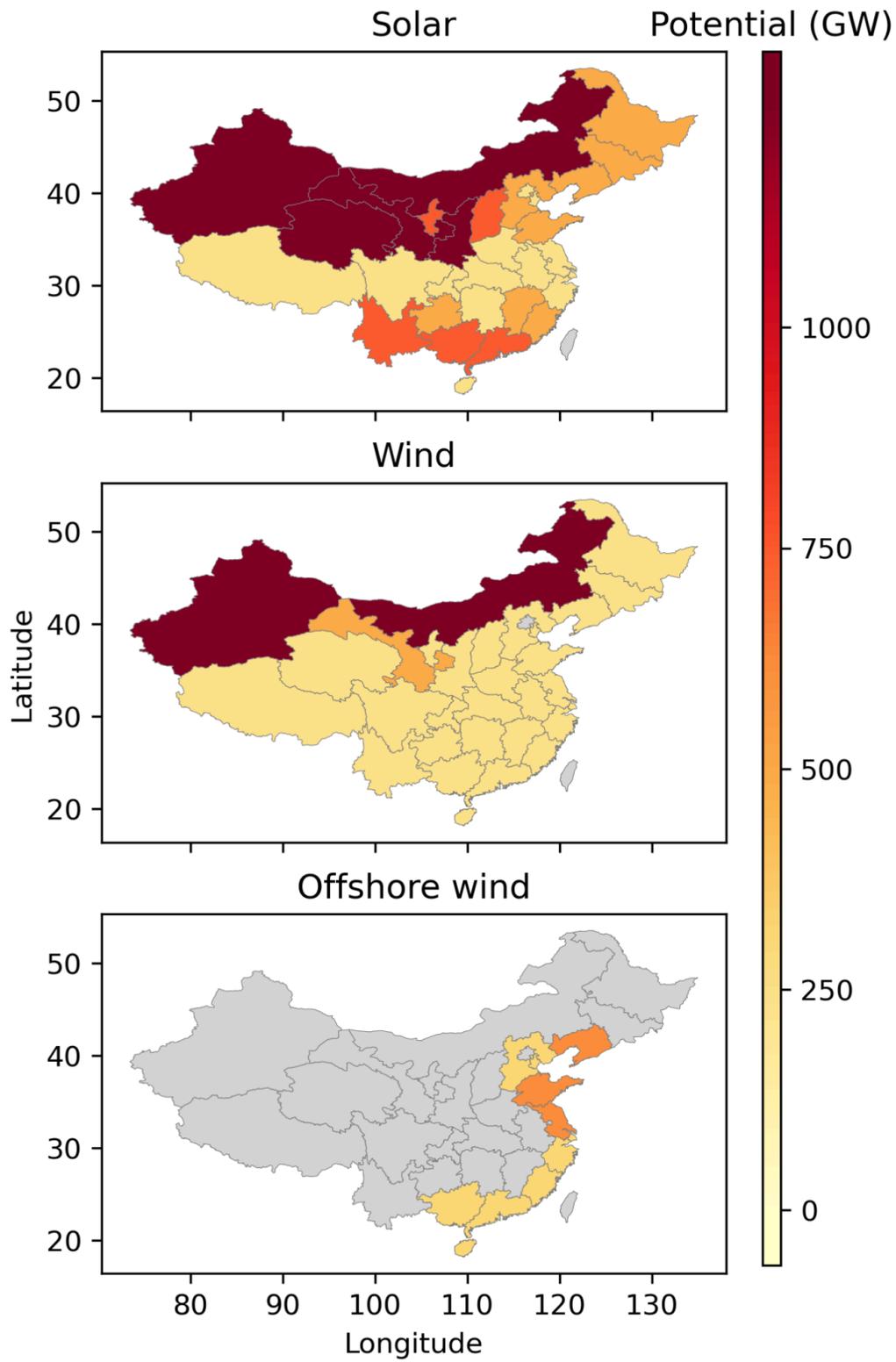

**Figure S5. The potential of solar, onshore wind and offshore capacities.**

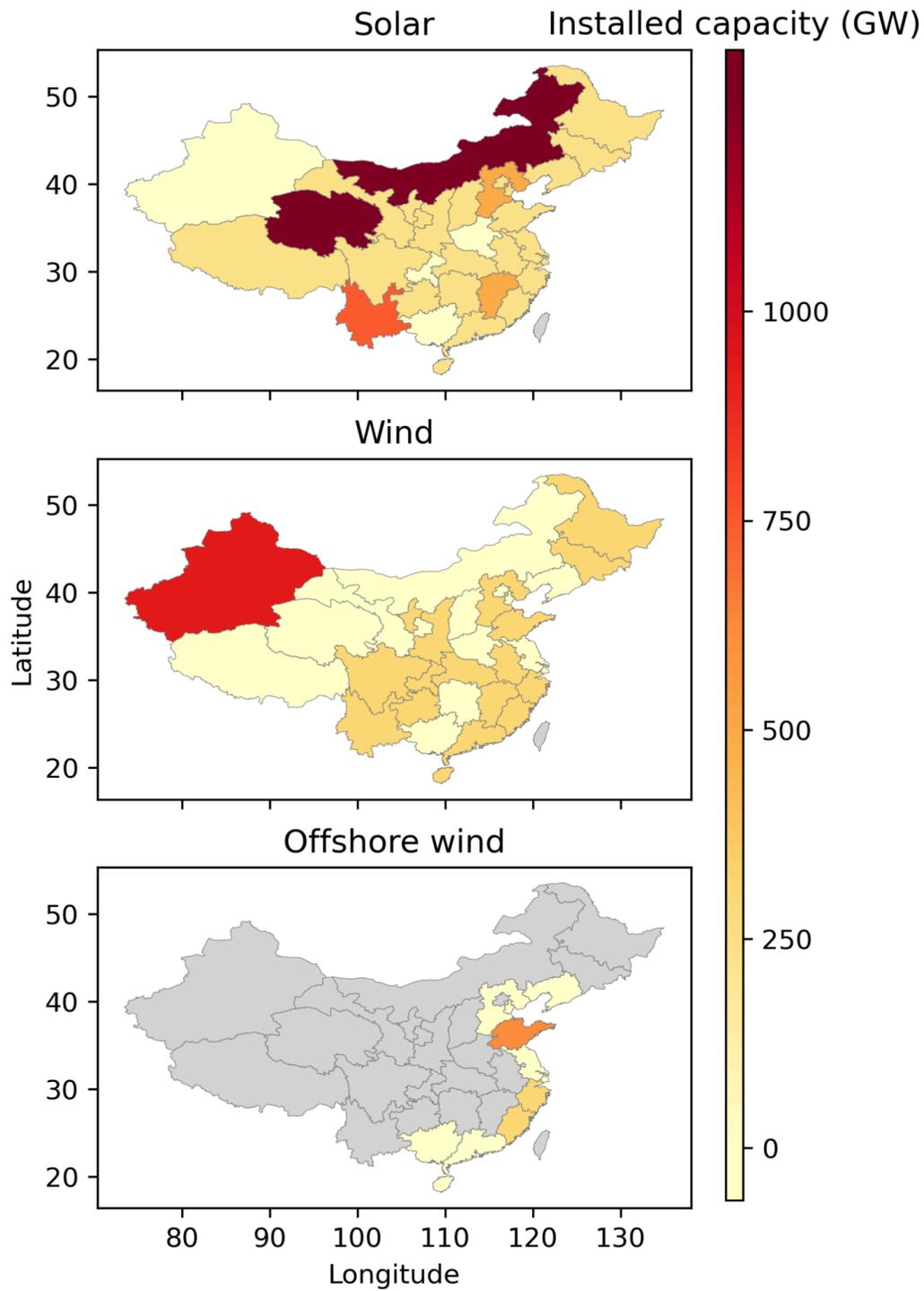

**Figure S6. The installed capacities of solar, onshore wind and offshore capacities under the ZE scenario.**

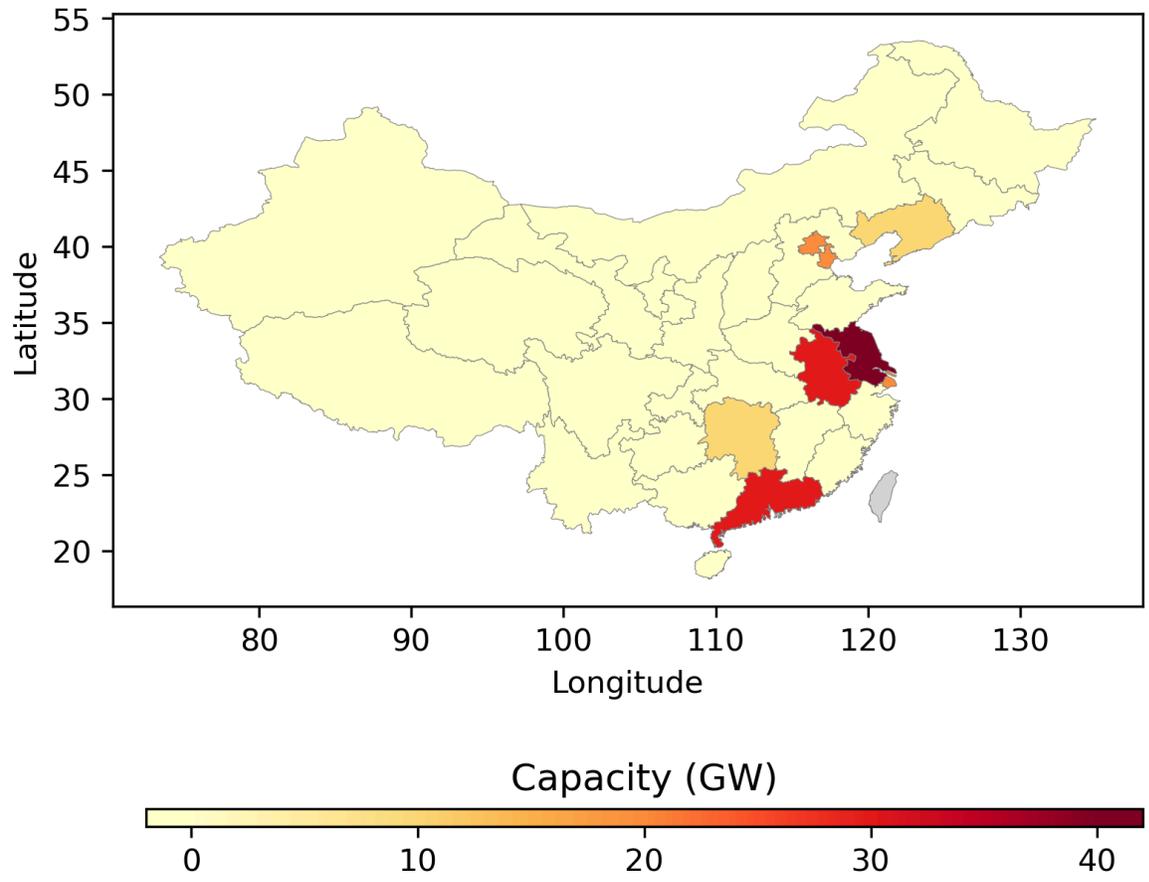

**Figure S7. Fuel cell capacities under the ZE scenario but without new transmission and hydrogen pipelines.**

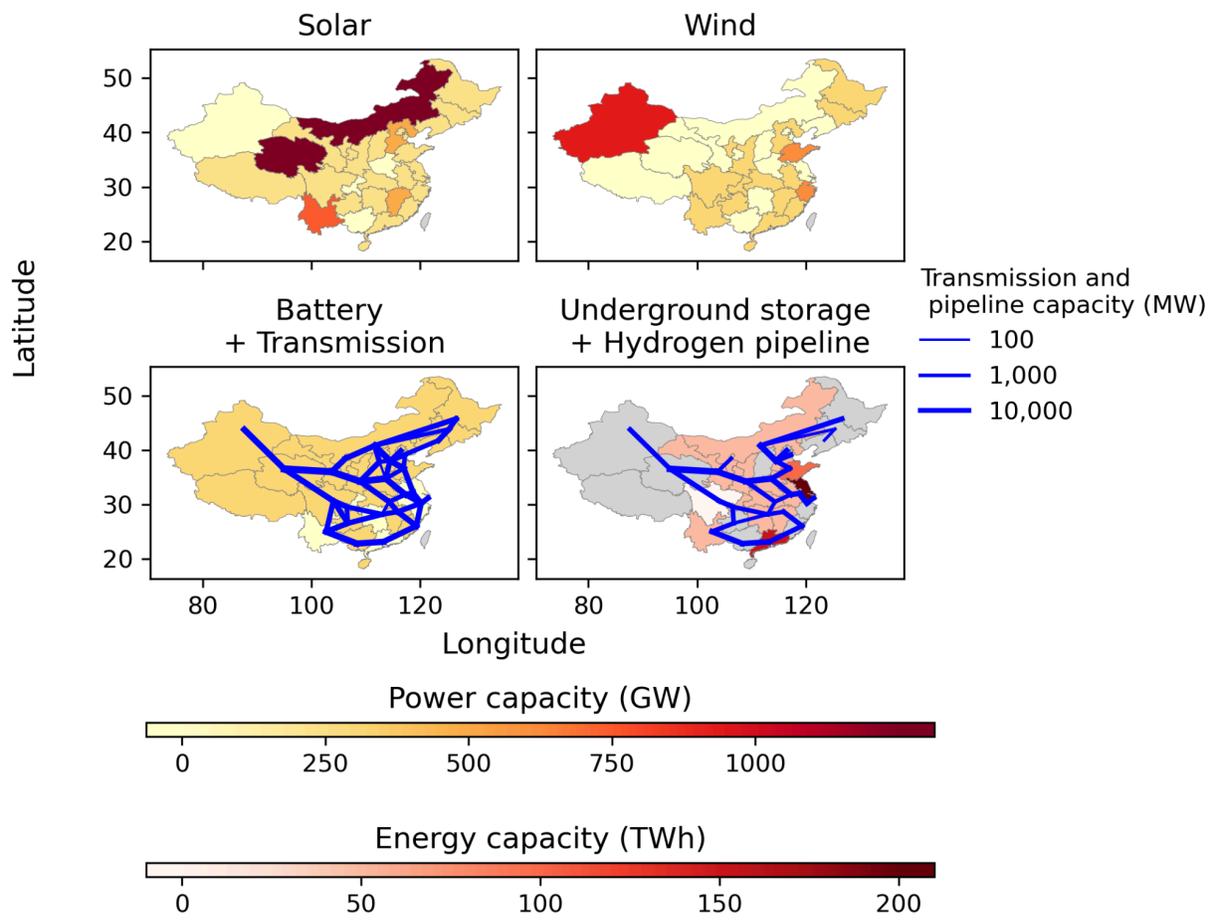

**Figure S8. Capacities of solar, wind, battery, underground storage, transmission lines and hydrogen pipelines. The color gray means that no underground storage is available in the area.**

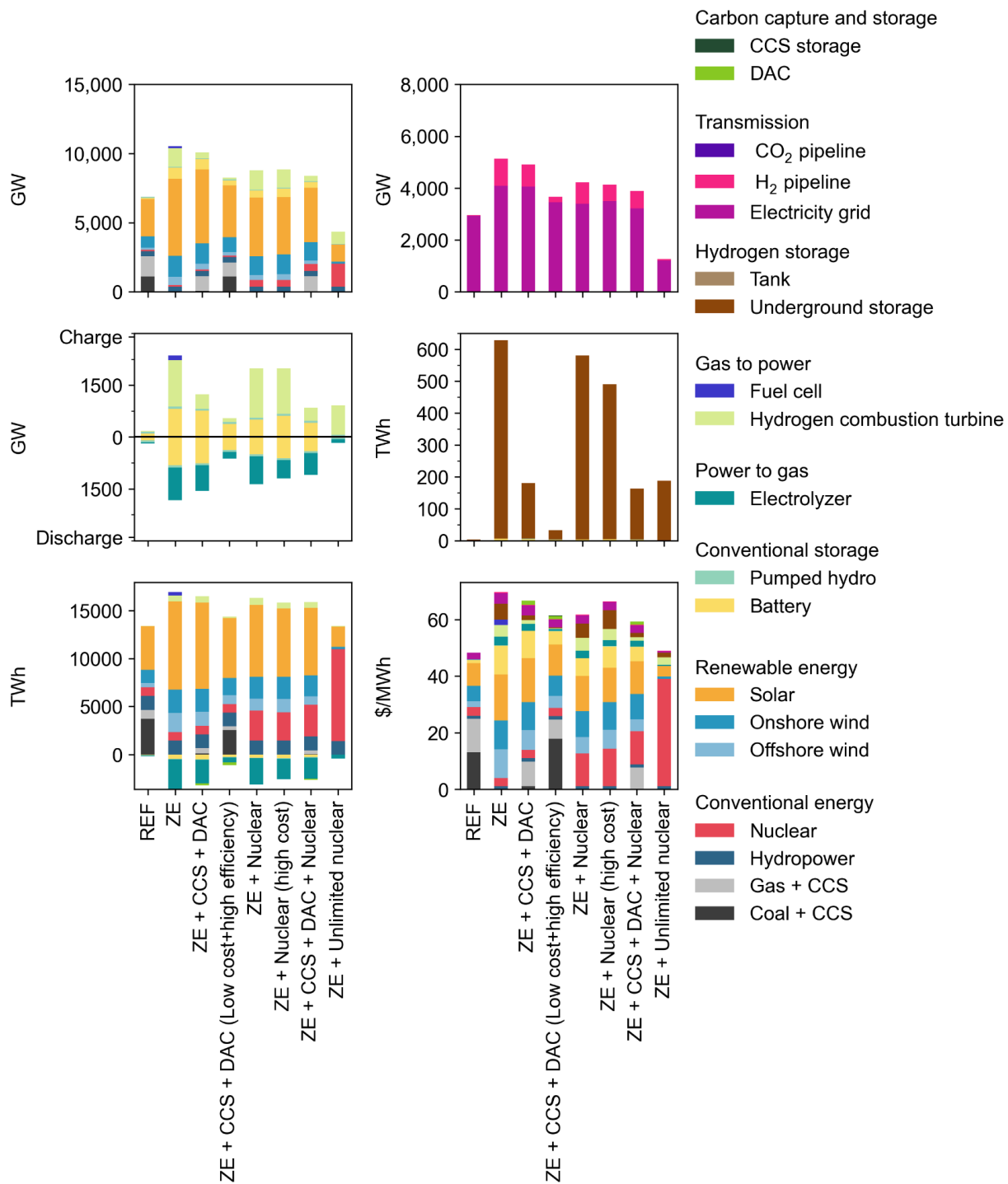

**Figure S9. a**. Generation capacity, **b**. transmission capacity, **c**. power capacity of charging and discharging storage, **d**. energy capacity of storage, **e**. generation, and **f**. levelized cost of electricity under different zero-carbon scenarios. Different colors refer to different technologies. REF refers to the scenario without a carbon emission target. The scenario 'CCS + DAC + Low

cost' refers to a scenario where the CCS and DAC follow the cost projection in 2050, and the CCS capture rate reaches 95%. 'Unlimited nuclear' refers to a scenario where no capacity cap is imposed on nuclear power plants.

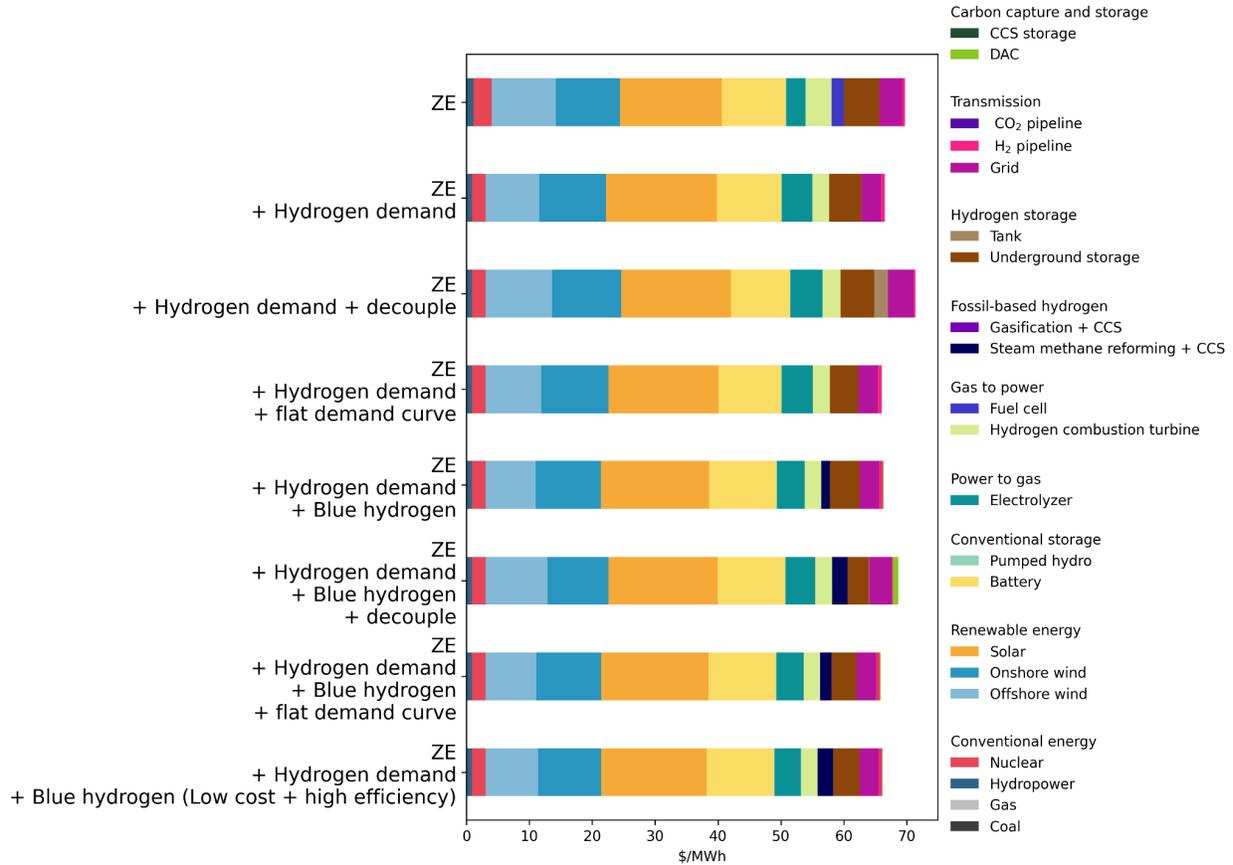

**Figure S10. Levelized cost of energy under different scenarios.**

The 'flat demand curve' refers to the scenario where in each province, the hourly hydrogen demand is the same across every hour. 'Blue hydrogen (Low cost + high efficiency)' refers to the scenario where the CCS and DAC follow the cost projection in 2050, and the CCS capture rate reaches 95%. 'decouple' refers to the scenarios where hydrogen production in the electricity system and hard-to-abate sectors is decoupled.

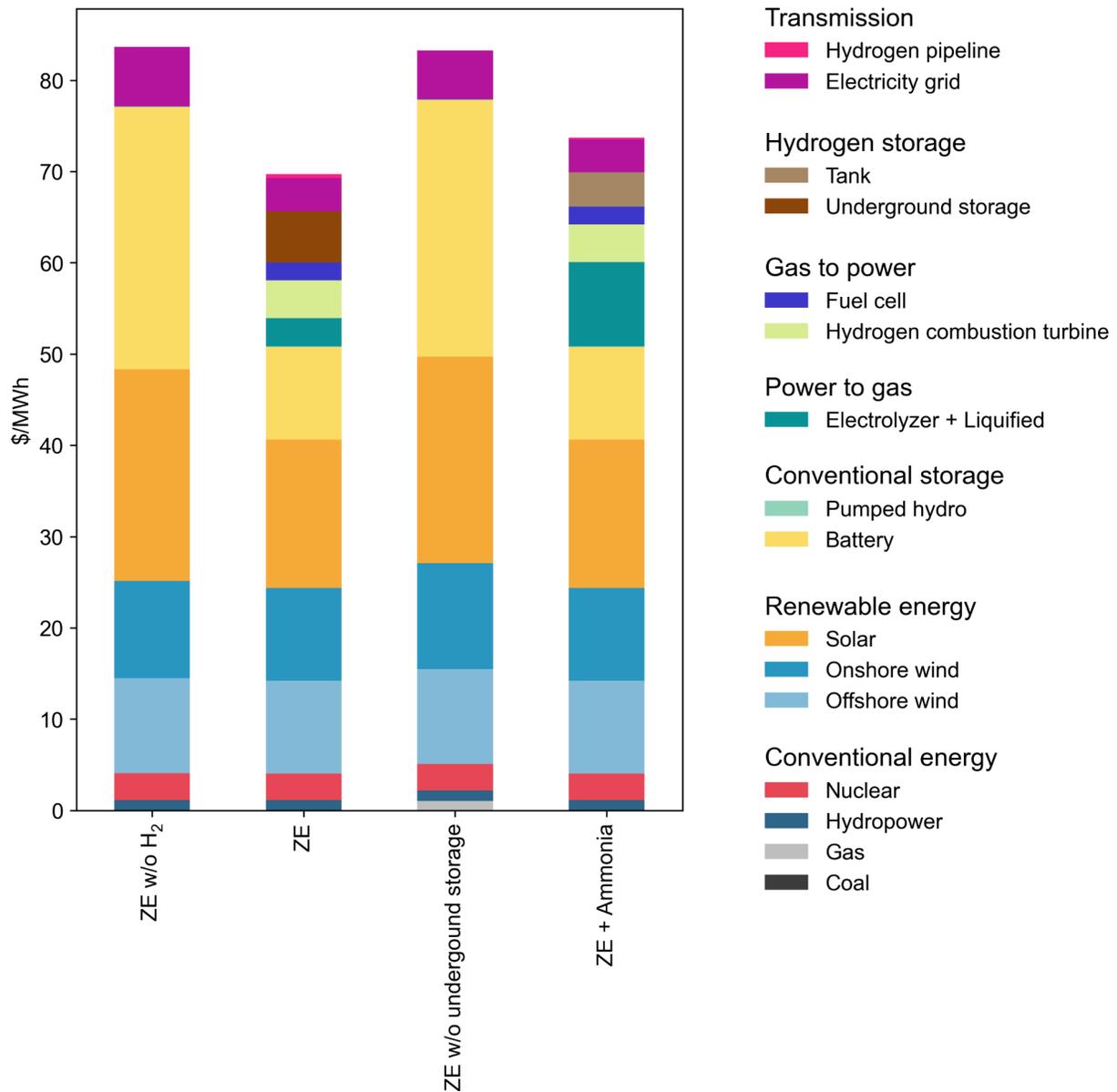

**Figure S11. Levelized cost of energy using ammonia.**

The ZE + Ammonia assumes the same capacity mix as ZE, but changes the costs of power-to-gas, hydrogen pipeline, and storage. By including the hydrogen-to-ammonia-to-hydrogen process, the cost of power-to-gas capacity is three times of the electrolyzer, the cost of transport is one-third of the hydrogen pipeline, and the cost of ammonia storage is two-thirds of the hydrogen underground storage. The cost of hydrogen-to-ammonia was collected from the Danish Energy Agency[5], and the cost of ammonia cracking, storage and transportation was collected from Argonne National Laboratory[13].